\documentclass[useAMS,fleqn,usenatbib]{mnras}  %fisso

\usepackage{newtxtext,newtxmath}
\usepackage[T1]{fontenc}
\usepackage{ae,aecompl}
\usepackage{graphicx}	% Including figure files
\usepackage{amsmath}	% Advanced maths commands
\usepackage{amssymb}	% Extra maths symbols
\input psfig.sty       % epsf.sty
\RequirePackage{xcolor}

\def \grss {GRS 1915+105~}
\def \grs {GRS 1915+105}

\def \rx {{\it RXTE}}
\def \rxx {{\it RXTE}~}

\defcitealias{Massaro2020a}{Paper~I}
\defcitealias{Massaro2020b}{Paper~II}

\title[A mathematical model for GRS~1915+105 - III]
{A non-linear mathematical model for the X-ray variability of the microquasar \grs~ - III: 
Low-frequency Quasi Periodic Oscillations} 

\author[E. Massaro, F. Capitanio, M. Feroci, T. Mineo]  
 {E. Massaro$^{1}$,
  F. Capitanio$^{1}$\thanks{E-mail:\texttt{fiamma.capitanio@inaf.it (FC);  marco.feroci@inaf.it (MF); teresa.mineo@inaf.it (TM)}},
  M. Feroci$^{1}$\footnotemark[1],
  T. Mineo$^{2}$\footnotemark[1] \\
 $^1$ INAF, IAPS, via del Fosso del Cavaliere 100, I-00113 Roma, Italy \\
 $^2$ INAF, IASF Palermo, via U. La Malfa 153, I-90146 Palermo, Italy \\
}
      
\date{ VERSIONE  3.2 ~~~ Roma 23 Giugno 2020 }

\pubyear{2020}

\begin{document}

\label{firstpage}
\pagerange{\pageref{firstpage}--\pageref{lastpage}}
\maketitle

% Abstract of the paper (max 250 words)   testo di 234 parole
\begin{abstract}
The X-ray emission from the microquasar \grss shows, together with a very complex 
variability on different time scales, the presence of low-frequency quasi periodic 
oscillations (LFQPO) at frequencies lower than $\sim$30 Hz.
In this paper, we demonstrate that these oscillations can be consistently and naturally 
obtained as solutions of a system of two ordinary differential equations that is able to 
reproduce almost all variability classes of \grs.
We modified the Hindmarsh-Rose model and obtained a system with two dynamical variables 
$x(t)$, $y(t)$, where the first one represents the X-ray flux from the source, and 
an input function $J(t)$, whose mean level $J_0$ and its time evolution is responsible 
of the variability class. 
We found that for values of $J_0$ around the boundary between the unstable and the 
stable interval, where the equilibrium points are of spiral type, one obtain an 
oscillating behaviour in the model light curve similar to the observed ones with a 
broad Lorentzian feature in the power density spectrum and, occasionally, with one 
or two harmonics. 
Rapid fluctuations of $J(t)$, as those originating from turbulence, stabilize the 
low-frequency quasi periodic oscillations resulting in a slowly amplitude modulated
pattern.
To validate the model we compared the results with real \rxx data which resulted 
remarkably similar to those obtained from the mathematical model.
Our results allow us to favour an intrinsic hypothesis on the origin of LFQPOs in 
accretion discs ultimately related to the same mechanism responsible for the spiking 
limit cycle.
\end{abstract}

\begin{keywords}  black hole physics – binaries: close – stars: individual: GRS 1915+105 – X-rays:stars.
\end{keywords}

\section{Introduction} %%%%%%%%%%%%%%%%%%%%%%%%%%%%%%%%%%%%%%%%    <<<<  SECTION  1
\label{sct:intro}

\begin{figure*}
\centering
\includegraphics[width=9.4cm,angle=-90]{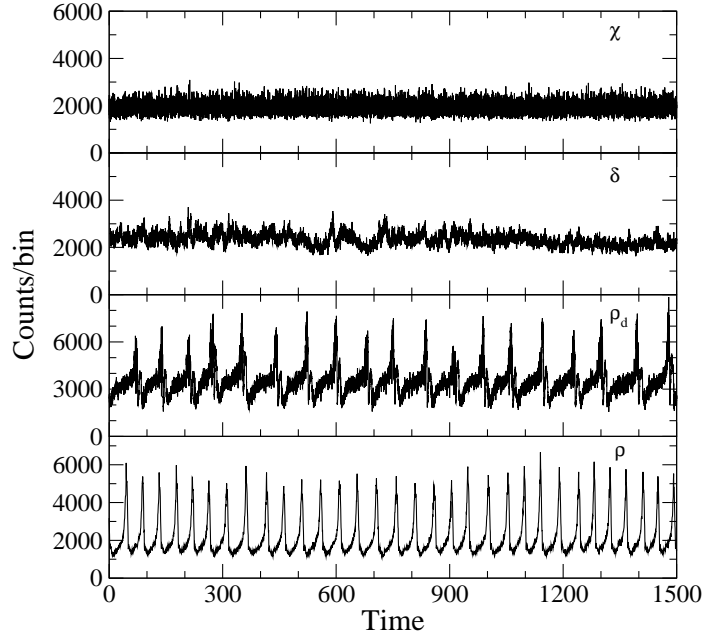}
\caption{
Examples of four types of \rxx light curves of \grs.
The greek letters correspond to the variability classes according to the definition
by \citet{Belloni2000}.
}
\label{fig:rxte_1}
\end{figure*}

\begin{figure*}
\centering
\includegraphics[width=9.4cm,angle=-90]{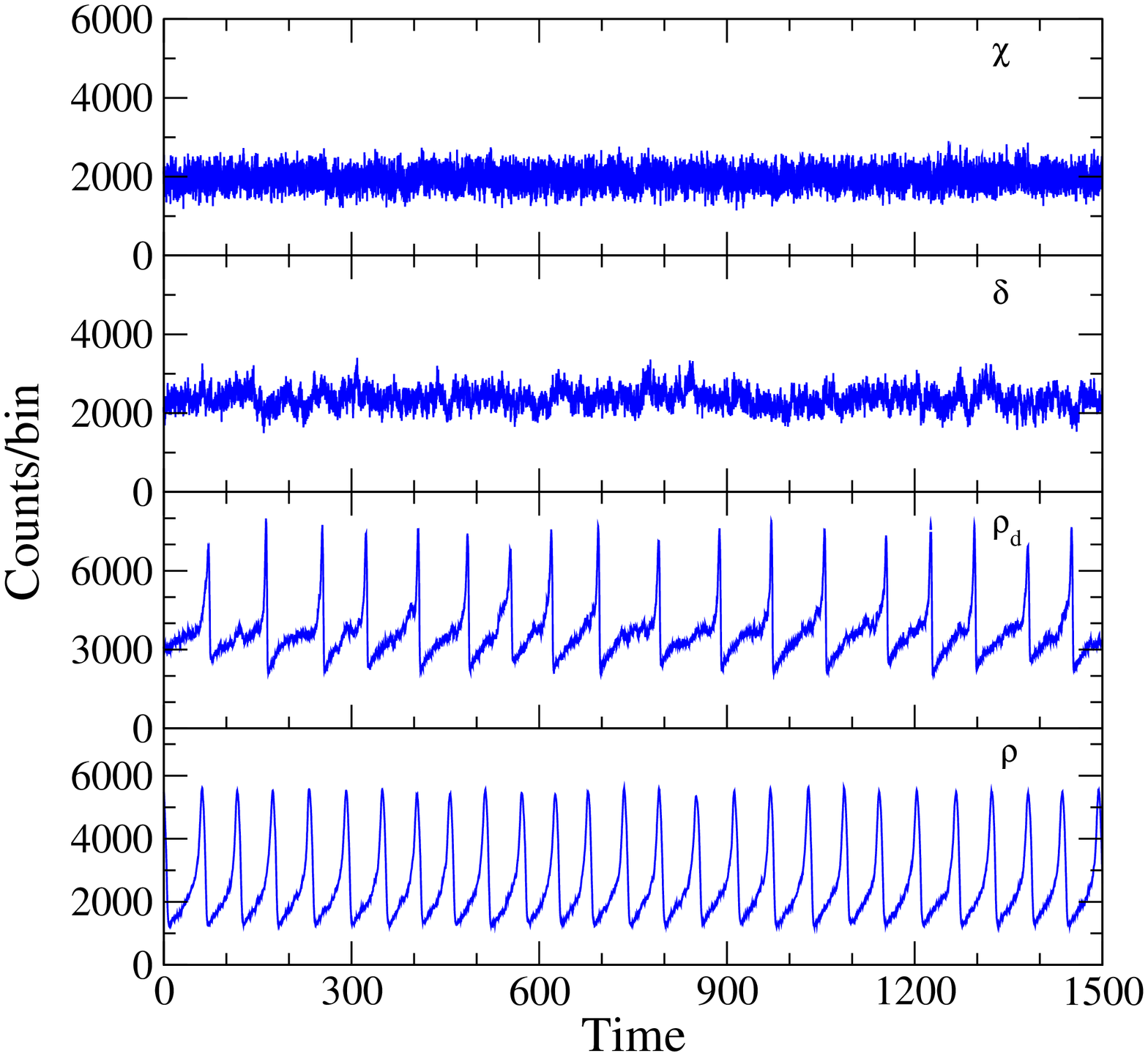}
\caption{
Solutions of the mathematical model described by \citetalias[][]{Massaro2020a} 
and \citetalias[][]{Massaro2020b} and reproducing the observed light curves \grss 
in Fig.~\ref{fig:rxte_1}.
These series, from top to bottom, are obtained by increasing only a parameter as
explained in Sect.~\ref{sct:ode}.}
\label{fig:mhr_1}
\end{figure*}

Quasi Periodic Oscillations (QPO) in X-ray binaries were discovered in the eighties
\citep{vanderKlis1989} and they were after detected in several Black Hole candidates 
(BHC) \citep{Remillard2006}.
Low-Frequency QPOs (LFQPO) are observed as broad peaks with a well approximate 
Lorentzian profiles in the Power Spectral Density (PDS) and centered at a frequency 
$\nu_0 \lesssim 30$ Hz  \citep[see the review by][]{Motta2016}.
LFQPOs have been classified in different types \citep{Wijnands1999, Remillard2006, 
Casella2004, Casella2005, Motta2016} according to the peak frequency and width, 
the relevance of the harmonics and the shape of the noise in the PDS.
Many observational studies have shown that central frequencies vary and exhibit
correlations with the mean brightness and with the energy of the photons.
Following \citet{vandenEijnden2016} one can classify the 
explanations of LFQPOs into the broad types of {\it geometric} and {\it intrinsic} 
models:
In the former class,  the source luminosity is not time modulated but has an anisotropic angular pattern 
and the flux oscillations are produced by changes of orientation with respect to the 
of sight line (e.g. precession or Lense-Thirring effect as proposed by \citet{Ingram2009}),                                                          
while in the latter type, the QPO origin is related to emissivity changes due to shocks 
\citep{Chakrabarti1993}, or variations of the accretion rate originated in various
kinds of instabilities \citep{Chen1992, Chen1995, Tagger1999, Varniere2012, Marcel2020}.

In the present paper we consider the well studied microquasar \grs, discovered by 
\cite{CastroTirado1992}.
This source exhibits a bright and highly variable X-ray emission, characterized by 
several different variability patterns, that alternate steady and noisy emission to 
regular and chaotic series of bursts.  
A first classification of the observed multifarious time behavior based on the 
signal structure and on the photon energy distribution was presented by \citet{Belloni2000}, 
who defined twelve classes identified by a greek letter, but 
new other patterns were observed on subsequent occasions. 
Some examples of X-ray light curves, from the \rxx data archive of four of these
variability classes are shown in Fig.~\ref{fig:rxte_1}.
In the top panel there is a typical $\chi$ class light curve that consists
of a rather steady and highly noisy signal with a nearly constant mean value, 
and in the bottom panel there is a $\rho$ class light curve with a long sequence 
of nearly regular bursts.
In the two intermediate panels there are examples of $\delta$ and $\rho_d$ signals,
the latter defined by \citet[][]{Massaro2020a}, that are considered transition
classes between stable and unstable states. 
A bursting $\rho$ class light curve was first reported by \citet{Taam1997}, 
who interpreted it as an evidence of a limit cycle in an accretion disc around 
a black hole originating from thermal-viscous instabilities 
\citep[see also][]{Taam1984, Szuszkiewicz1998}.

LFQPOs are frequently observed also in \grss \citep{Paul1997, Fender2004}.
\citet{Markwardt1999} and \citet{Muno1999} found that these oscillations 
occur during the dips, when the source is in a flaring state, and that their frequency 
correlates with the parameters of the thermal disc component, like the temperature.  
\citet{Rodriguez2002} confirmed that frequency variations are well correlated 
with the soft X-ray flux and proposed that they could be related to a hot point 
in an optically thick disc, while the presence of harmonics could be a signature 
of a non-linear instability.
The high X-ray flux from \grss allows for detailed investigations on LFQPOs which
showed the existence of a modulation with the QPO phase either of the observed reflection 
fraction or of the iron line shape that change throughout the cycle \citep{Ingram2015}.

The stability of accretion discs is also a very interesting subject of 
investigations since many years and theoretical analysis suggested that 
thermal and viscous instabilities can develop and establish a limit cycle 
behavior.
The complex hydrodynamical, thermal and magnetic phenomena occurring in accretion
discs around black holes involve non-linear processes whose evolution are 
described by a system of partial differential equations, whose solutions
are obtained by numerical calculations involving several quantities not 
directly observable, as the gas density or viscous stresses.

In a recent couple of papers, \citep[][hereafter Paper I and Paper II]{Massaro2020a, Massaro2020b}, 
we showed that the solutions of a system of non-linear ordinary differential equations 
(ODE) reproduce several classes of the X-ray light curves of \grs.
This system, named Modified Hindmarsh-Rose (shorthly MHR), is a modified version of 
the well studied Hindmarsh-Rose model that is used for describing neuronal bursts.  
MHR model is a non-autonomous system with a time dependent input function.
The function we adopted has a variable component added to fast random fluctuations 
introduced to simulate a possible plasma turbulence in the emitting source.
Some examples of the numerical solutions obtained in \citetalias{Massaro2020a} are 
given in Fig.~\ref{fig:mhr_1}. 
These light curves, remarkably similar to the true data in Fig.~\ref{fig:rxte_1}, 
are obtained by changing the value of only one parameter.
It is interesting that the shape of one of the equilibrium curves of the 
proposed ODE system presents a S-pattern similar to those derived from numerical 
solutions of disc equations (see the review by \citet{Lasota2016}).
In stable states, like the $\chi$ class, that is much more frequently observed 
than all the other ones \cite{Belloni2000}, the X-ray flux  of \grss remains nearly 
constant, but a large noise component is present.
This component may be particularly relevant for reproducing some variability 
classes and can play a very important role in the origin of LFQPOs.

In the present paper we investigate in detail how the MHR model produces
LFQPOs and how the amplitude of the noisy component affects the characteristics 
of the solutions.
In particular, we present some results on LFQPOs in \grss applying a 
harmonic filtering method and, after a brief description of our mathematical model, 
we demonstrate that it can also account for LFQPOs with some features remarkably 
similar to the observed ones.
These findings can be applied to other sources as well.

\section{LFQPO analysis of \rxx observations} 
%%%%%%%%%%%%%%%%%%%%%%%%%%%%%%%%%%%    <<<<  SECTION  2
\label{sct:rxte_obs}

\begin{figure}
      \hspace{0cm}
%\epsfysize=12.0cm
%\epsfbox{A8phs.ps}
\includegraphics[height=8.5cm,angle=-90,scale=1.1]{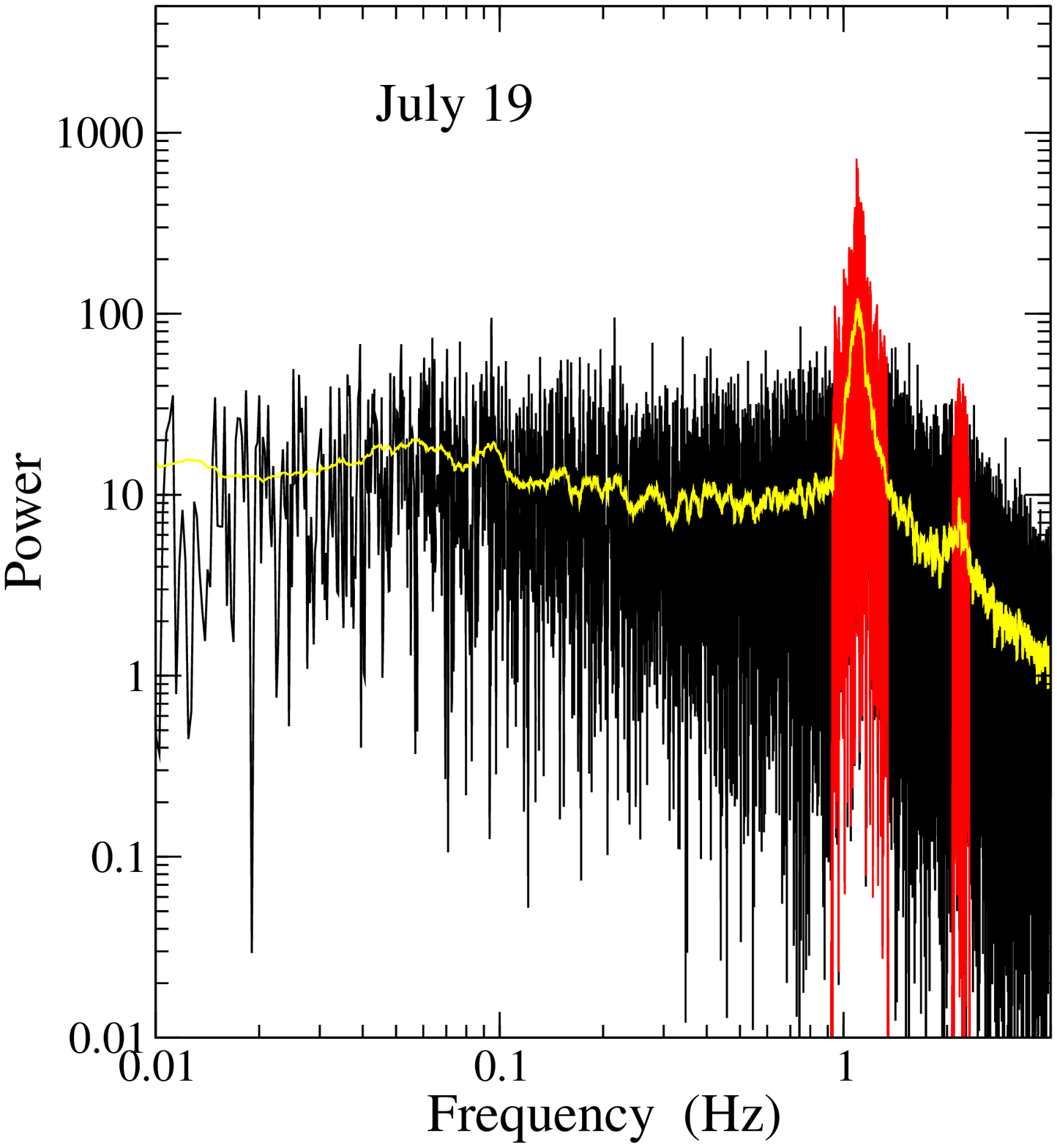}\\
\includegraphics[height=8.5cm,angle=-90,scale=1.1]{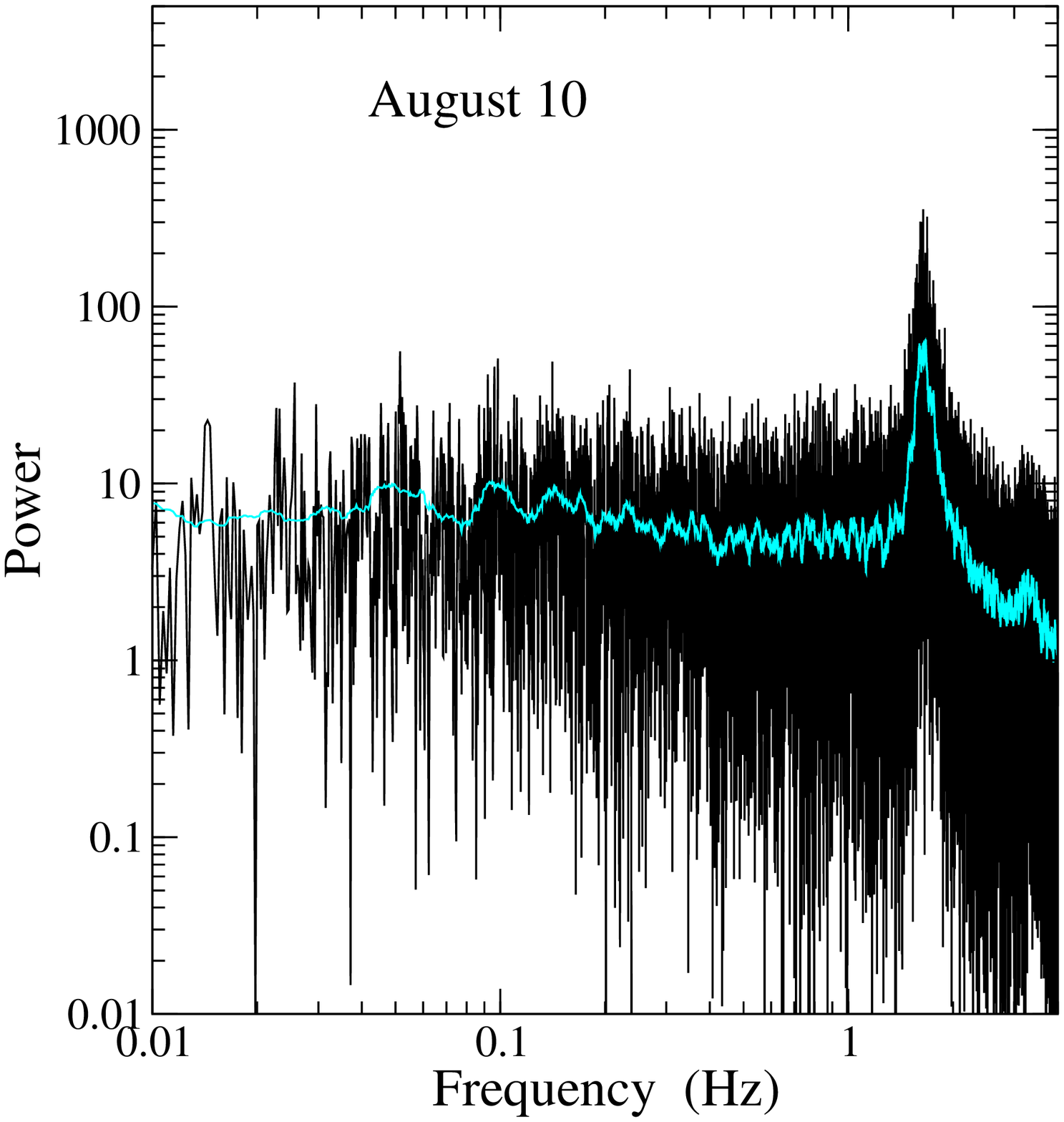}
\caption[]{
Upper plot: Fourier spectrum of the light curve of \grss of the \rxx observation 
on 1996 July 19 with a QPO broad peak centered at the frequency of 1.11 Hz.
the yellow data are after running average smoothing to reduce the noise.
Red data are those obtained after the two bandpass filtering on the fundamental
and 1st harmonic feature used for deriving the QPO signal.\\
Lower plot: Fourier spectrum of the light curve of \grss of the \rxx observation 
on 1996 August 10 with a QPO broad peak centered at the frequency of 1.66 Hz.
the cyan data are obtained  running average smoothing to reduce the noise.
}
\label{fig:rxte_qpo_pds}
\end{figure}

\begin{figure}
      \hspace{0cm}
%\epsfysize=12.0cm
%\epsfbox{A8phs.ps}
\includegraphics[height=7.5cm,angle=-90,scale=1.1]{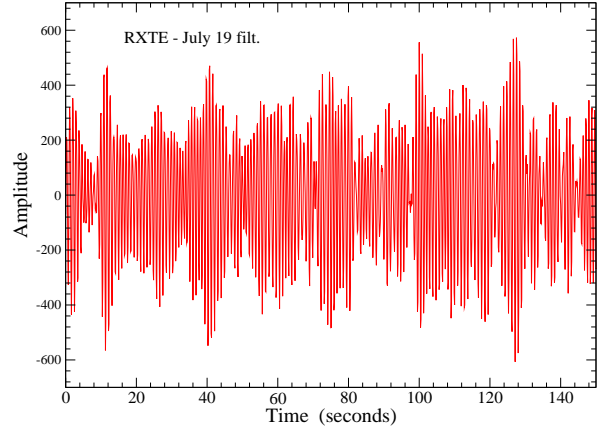}
\caption[]{ A short segment of the X-ray light curve of the \rxx observation 
of 1996 July 19 of \grss obtained by means of the two band Fourier filtering 
shown in the upper plot of Fig.~\ref{fig:rxte_qpo_pds}.
}
\label{fig:rxte_qpo_lc}
\end{figure}

Extensive investigations of several \rxx observations of \grss focused on QPO detection 
and their properties were performed by \citet{Morgan1997} and \citet{Yan2013}; 
more recently, \citet{vandenEijnden2016} reported new results on another sample 
of \rxx data sets.
We selected in these data a couple of observations having LFQPO features with high 
quality factors, and precisely those with ID 1040801-25-00 and 1040801-29-00a,  
both performed in 1996, the former on July 19 and the latter on August 10.
The \rx/PCA light curves were extracted in standard 1 mode, namely in the total energy 
band (2-40 keV) and with the time binning of 125 ms.
Both observations include more than a single orbit and we selected only the first one
to avoid time gaps in the data; then we computed their PDS by means of a standard 
Discrete Fourier transform algorithm.
The July 19 series has a duration of 3296 s and that of August 10 of 2816 s. 

In the two panels of Fig.~\ref{fig:rxte_qpo_pds}, we report the PDS of both observations,
whose values (black spectra) are affected by a very large noise that can be reduced by 
performing a simple running average smoothing; broad Lorentzian peaks centered at 1.11 Hz
(July 19) and 1.66 Hz (August 10), in a very good agreement with the above quoted papers,
are well apparent.
Note that in both spectra a first harmonic feature is also present.

We reconstruct the time signal corresponding to these features applying a technique 
similar to that used by \citet{vandenEijnden2016} that consists in  
filtering both the real and imaginary series of the discrete Fourier transform of the 
photon count rates.
We did not use the same optimal filter used by these authors but applied the simple 
rectangular bandpass with a smooth tapering at the boundaries given by the following 
formula:

\begin{eqnarray}
\Phi (X) &=& \frac{1}{4} \left(1+\frac{(X-X_1)}{\sqrt{(X-X_1)^2 + s_1^2}}\right) \times 
                      \nonumber  \\
         & & \left(1-\frac{(X-X_2)}{\sqrt{(X-X_2)^2 + s_2^2}}\right)
\label{eq:filter}
\end{eqnarray}

\noindent
where $X_1$ and $X_2$ are the two frequencies defining the accepted window and $s_1$ 
and $s_2$ rules the slopes for tapering the filter profile (for a symmetric filter
$s_1 = s_2$).  
In our case, we considered also the power in the first harmonic to better approximate
the true waveform.
The filtered PDS spectrum of the July 19 data is shown in red in Fig.~\ref{fig:rxte_qpo_pds}, 
showing that our method is not largely different from that used by \citet{vandenEijnden2016}.

\begin{figure}
\includegraphics[height=8.8cm,angle=-90,scale=1.0]{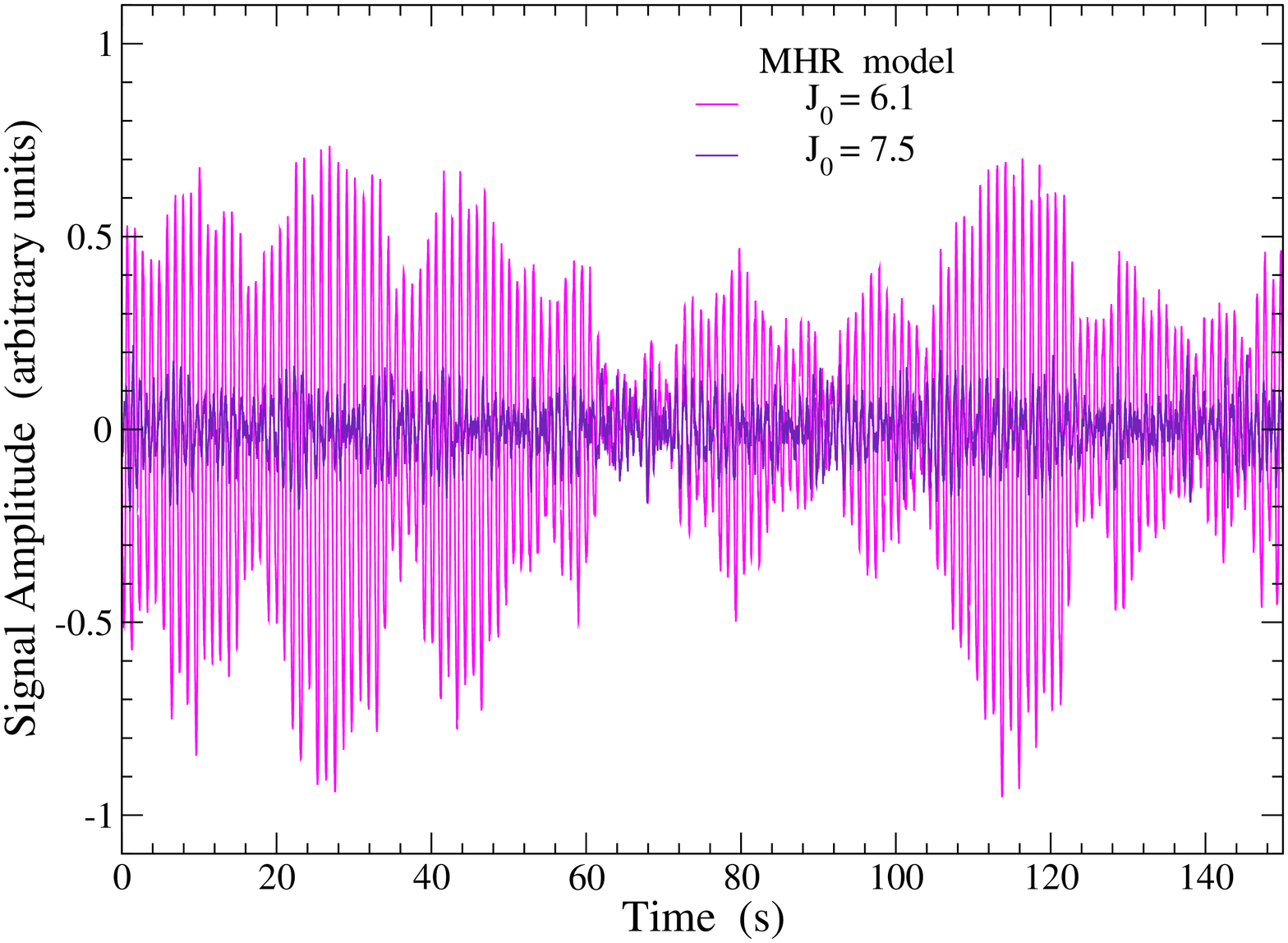}  \\
\includegraphics[height=8.8cm,angle=-90,scale=1.0]{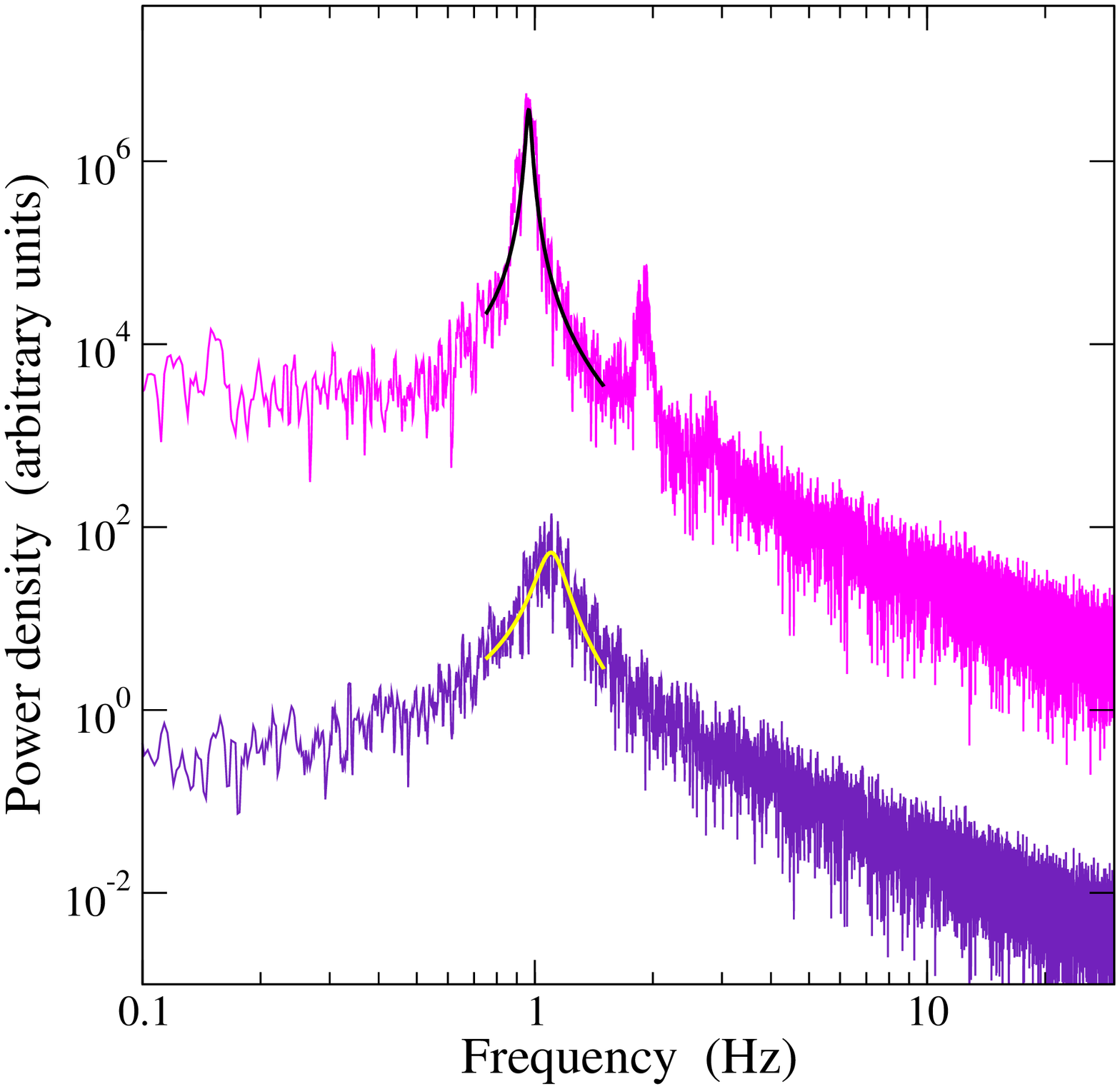} 
\caption[]{
Upper plot: Light curves computed in \citetalias{Massaro2020a} by means of the MHR 
model assuming $J(t)$ of equation~\ref{eq:partial}, with $C = 3.5$ and two values 
of $J_0 = 6.1$ (upper magenta spectrum), just higher than the transition level 
and 7.5 (lower indigo spectrum), both located in the stable region.
Lower plot: power density spectra of these curves, shifted to separate the profiles 
of QPO features.
The black and yellow thick curves are the Lorentzian fits to the QPO peaks.
}
\label{fig:cpsl}
\end{figure}

A segment of the signal obtained by the inverse Fourier transform is shown in 
Fig.~\ref{fig:rxte_qpo_lc}.
The signal structure presents an amplitude modulation applied to a carrier at the 
frequency of the central value of LFQPO peak like the waveform reported by 
\citet{vandenEijnden2016}.

As already shown in \citetalias{Massaro2020a}, the same MHR mathematical model used for 
computing the light curves reported in Fig.~\ref{fig:mhr_1} is also able to produce 
LFQPOs with only a further increase of the input parameter and without any changes of 
the other parameters.
In Fig.~\ref{fig:cpsl} we reported two series computed in the quoted paper and their 
PDS to show how the model gives LFQPOs remarkably similar to the true data.
Light curves show a modulated fast oscillation with an amplitude depending on the input
parameter.
Note also the large differences in the amplitudes of these signals which are due
to the  values of $J_0$ but not to the amplitude of the noise, as explained in the
next section.
In the lower panel, the upper spectrum has a well apparent second harmonic and  the 
third one is also marginally detectable, whereas in the lower spectrum only the peak at the 
fundamental frequency is visible.
In the same plot we report the Lorentzian best fits to the peaks, which are in a very
well agreement with their profiles and have quality factors $Q = \nu / \Delta \nu$ 
equal to 29.2 and 5.8, comparable to observed values.

Thus one can rise the hypothesis that the spikes of the limit cycle and LFQPO, both
frequently observed in \grs, have a common origin and their occurrence depends on the
value of only one parameter.
In the next section we summarize the main mathematical properties of the MHR model and 
extend the analysis of the nature of the equilibrium points in order to make clear the 
necessary conditions for developing LFQPOs.
Moreover, the MHR model will make possible to investigate the relevant role of the 
presence of a noise component in stabilizing LFQPOs, which, in absence of such a 
component, would be rapidly damped.

\section{The MHR non-linear ODE system} 
%%%%%%%%%%%%%%%%%%%%%%%%%%%%%%%%%%%%%%%%    <<<<  SECTION  3
\label{sct:ode}

As stated in the introduction, in \citetalias{Massaro2020a} and \citetalias{Massaro2020b} 
we reproduced the rich and complex behavior of \grss by means of a non-linear system 
of ODE as those used for describing quiescent and bursting signals in neuronal arrays.
This approach offers the possibility of describing transitions between stable 
and unstable equilibrium states with the onset of limit cycles.
The original Hindmarsh-Rose model (see the historic review of \citealt{Hindmarsh2005}, 
and the tutorial paper \citealt{Shilnikov2008}) was based on three ODEs, for three
dynamical variables $x$, $y$ and $z$, involving changes on different time scales. 
In our previous works \citepalias{Massaro2020a, Massaro2020b}, we considered a modified 
system without the variable $z$ and including an external  input function of the 
time $J(t)$.
Moreover, we adopted the simplifying assumption of taking the same quadratic 
coefficient in both equations, and without loss of generality, the cubic coefficient
was assumed equal to 1.0.
The resulting modified system, therefore, is non autonomous and includes only two 
ODEs that, using the same notation as in \citetalias{Massaro2020a}, are:

\begin{equation}
\left\{
\begin{array}{lcl}
\dot{x} &=& - x^3 + \beta x^2 + y + J(t) ~~ = ~~ f(x,y)\\ 
& & \\
\dot{y} &=& - \beta x^2 - y  ~~~~ = ~~ g(x,y)
\end{array}\right.
\label{eq:ode}
\end{equation}

\noindent
where the signs of the various terms were taken to have positive parameters' values. 
As in our previous papers we consider only the $x$ time series that represents
the X-ray photon flux of the source.
Of course the solutions must be scaled both in time and amplitude to be compared
with the observational data.

\subsection{Nullclines, equilibrium points and stability for a constant input} 
%%%%%%%%  <<<<  SUBSECTION  
\label{sct:nullc}

In the simple case of a constant $J(t) = J_0$, an assumption that makes the model  
autonomous, the equilibrium conditions of equation~\ref{eq:ode}, i.e. 
$\dot{x} = \dot{y} = 0$, are:

\begin{equation}
 y =  x^3 - \beta~ x^2 - J_0  
\label{eq:ode_eq1}
\end{equation}

\begin{equation}
 y = - \beta~ x^2  
\label{eq:ode_eq2}
\end{equation}

\begin{figure}
      \hspace{0cm}
%\epsfysize=12.0cm
%\epsfbox{A8phs.ps}
\includegraphics[height=7.5cm,angle=-90,scale=1.0]{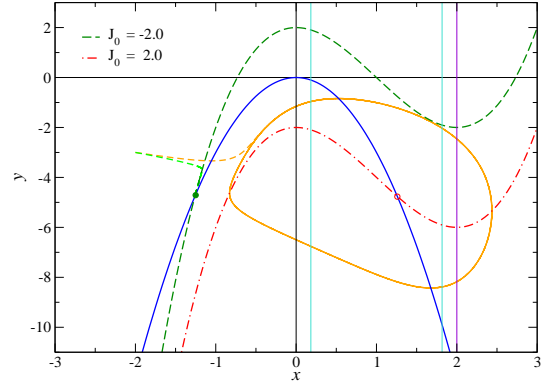}
\caption[]{
Nullclines for the MHR model with $\beta = 3$ and for two $J_0$ values 
equal to $-2$ (green long dashed curve) and $2$ (red dot-dashed curve); the solid blue 
curve is the parabola given by equation~\ref{eq:ode_eq2}.
Two phase trajectories (short dashed lines), one stable and the other describing a 
limit cycle, with equal initial conditions and two $J_0$ values are shown; green filled 
circle and red open circle are their equilibrium points.
The turquoise vertical lines delimit the unstable equilibrium interval, and the violet
line corresponds to the minima of cubic nullclines.
}
\label{fig:ncl}
\end{figure}

\noindent
The system admits only the real solution $x_* = J_0^{1/3}$, $y_* = - \beta ~J_0^{2/3}$,
that corresponds to the equilibrium point.
In Fig.~\ref{fig:ncl} we plotted in the plane $x,y$ the curves of equations~
\ref{eq:ode_eq1}~\ref{eq:ode_eq2}, 
named nullclines, for $\beta = 3.0$ as in \citetalias{Massaro2020a}: they intersect 
at the equilibrium point, that, as we demonstrated in that paper, it results 
always stable for $J_0 < 0$ while there is only one unstable interval for $J_0 > 0$.

\begin{figure}
      \hspace{0cm}
%\epsfysize=12.0cm
%\epsfbox{A8phs.ps}
\includegraphics[height=8.0cm,angle=-90,scale=1.1]{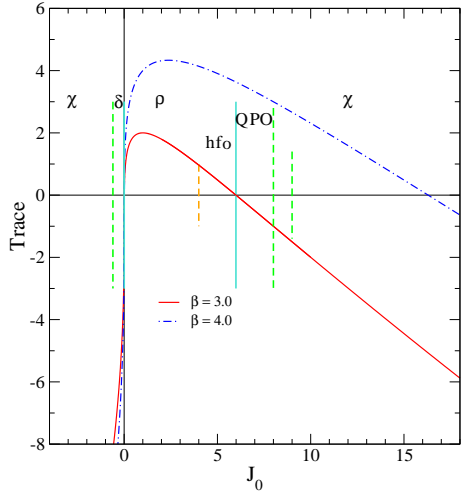}
\caption[]{
Plot of the traces of the Jacobian for two values of $\beta$ as function 
of the mean input $J_0$.
Stable  intervals ($\chi$, QPO) for the curve relative to $\beta=3.0$ (solid red line)
are outside the long vertical turquoise solid segments, while the unstable interval 
($\rho$) is between them.
The QPO interval is indicated in the stable interval between the higher stability
boundary and the local minimum of the cubic nullcline.
The `high frequency oscillation' (hfo) in the unstable interval is also indicated.
Dashed vertical lines within the stable and unstable regions mark the intervals in 
which other variability classes and behavior are obtained as reported in the graph.
}
\label{fig:tr}
\end{figure}

As shown in \citetalias{Massaro2020a}, the nature of the equilibrium depends only upon 
the sign of the trace of the Jacobian of the system 

\begin{displaymath}
\left(  \begin{array}{cc}
-3 x_*^2 +2 \beta x_* & 1  \\
-2 \beta x_* & -1  
\end{array} \right)
\label{eq:jaco}
\end{displaymath}

\noindent
evaluated at $(x_*, y_*)$, because the determinant is always non negative.
The zeroes of the trace 

\begin{equation}
 x_{1,2} = \frac{1}{3} (\beta \pm \sqrt{\beta^2 - 3})
\label{eq:zero}
\end{equation}

\noindent
define an interval within the trace is positive and therefore the equilibrium is
unstable.
For $\beta = 3.0$, this unstable interval for the variable $x$ is [0.1835, 1.8165] and 
is entirely contained in the interval [0.0, 2.0] that corresponds to the portion of 
the nullclines between the local maximum and minimum where the slope is negative,  
as it is easy to verify from the roots of the $x$ derivative of equation~\ref{eq:ode_eq1}.
It is important to note that the instability interval on $x$ depends only 
upon $\beta$ but not upon $J_0$; thus a change of this parameter moves the location 
of the equilibrium point allowing transitions between stable and unstable states.
However, we can relate the stability to $J_0$ computing the values of the trace  
when $J_0 = x_*^3$ varies; the resulting curves, for $\beta$, 3.0 and 4.0, are 
reported in Fig.~\ref{fig:tr}.
The two previous limits define three intervals for $J_0$, that we indicate as 
$S_1 \equiv [-\infty, 0.0061792...]$, $I \equiv [0.0061792..., 5.99382...]$ and 
$S_2 \equiv [5.99382..., +\infty]$.
In this figure, the two turquoise vertical solid lines delimit the unstable interval $I$ for the 
former value of $\beta$ given above, while the equilibrium in the intervals 
$S_1$ and $S_2$ is stable, but the trajectories in the phase space approaching 
to this state are different as explained in Sect.~\ref{sct:stable_solut_Wn}.
When $J_0$ varies slowly across the limits of $I$, transitions from stable to 
unstable equilibrium and {\it viceversa} occurs thus ruling the onset 
or the disappearance of the limit cycle.

Examples of stable and unstable dynamical solutions are also illustrated in Fig.~\ref{fig:ncl}, 
where two trajectories in the phase space corresponding to the values of $(x(t)$, $y(t))$
of the system in equation~\ref{eq:ode} are also plotted:
they start from the same initial position, $x_0 = -2.0$, $y_0 = -2.0$, but, while 
the one for $J_0 = -2.0$ reaches the green dashed nullcline and then moves directly 
towards the corresponding equilibrium point; the other (orange trajectory), computed
fixing $J_0 = 2.0$, crosses the parabolic nullcline and evolves to a closed orbit 
({\it limit cycle}) around the unstable (red open circle) equilibrium point.

\section{Stable solutions: numerical results} 
%%%%%%%%%%%%%%%%%%%%%%%%%%%%%%%%%%%%%%%%    <<<<  SECTION 4
\label{sct:stable_solut}

Our first step was the computation of some light curves in the case of a stable equilibrium.
We consider first the condition $J(t) = J_0$ without any noisy component that are useful
for describing the nature of equilibrium points and the evolution of phase space trajectories;
subsequently, we will present the results when random fluctuations are included in $J(t)$.
In all the following calculations we will assume $\beta =3.0$, as in \citetalias{Massaro2020a}.
In the study of LFQPOs, the stable solutions for negative values of $J_0$ are not interesting
and therefore we focus on the case of positive values of this parameter.
Numerical computations were performed by means of a Runge-Kutta fourth order integration 
routine \citep{Press2007}.

\begin{figure}
      \hspace{0cm}
%\epsfysize=12.0cm
%\epsfbox{A8phs.ps}
\includegraphics[height=7.5cm,angle=-90,scale=1.1]{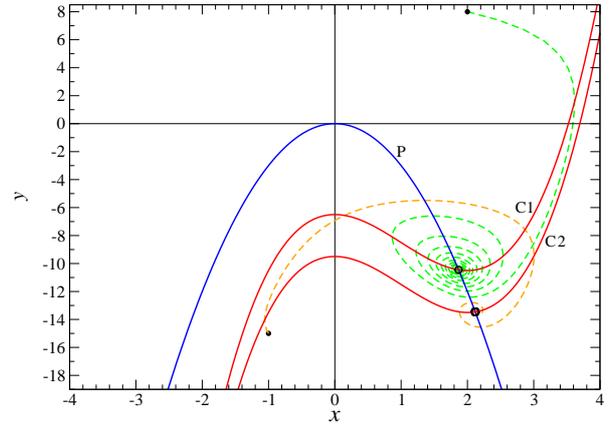}
\caption[]{
Nullclines for the system of equation~\ref{eq:ode} with $\beta = 3$ and for two $J_0$ values 
equal to $6.5$ (red curve C1) and $9.5$ (red curve C2); the blue (curve P) is the 
parabola given by equation~\ref{eq:ode_eq2}.
Stable equilibrium points are at the intersections of the parabola with the cubic lines
and are marked by open circles.
Two phase trajectories (dashed lines) with the different initial points, marked by black
solid circles, spiralling towards the equilibrium points are also reported.
}
\label{fig:ncspi}
\end{figure}

\subsection{Solutions without noise}          
%%%%%%%%%%%%%%%%%%%%%%%    <<<<  SUBSECTION
\label{sct:stable_solut_Wn}                       %%%      >>>>>>>>>>>>  REFANS Pt.4

The $x, y$ trajectories in the phase space have a rather simple pattern 
with a transient phase, depending on the initial values, followed by a rapid approach 
to a steady condition (equilibrium or limit cycle).
In Fig.~\ref{fig:ncl} the trajectory for $J_0 = -2.0$, after the initial transient, moves 
very close to the cubic nullcline and then it reachs the equilibrium point.
This behavior is always observed for all values of $J_0 \in S_1$.
Stable trajectories for $J_0 \in S_2$ have again an initial transient, but when the 
point is close to the cubic nullcline starts to describe a spiral that approximates 
an elliptical shape when its amplitude decreases converging to the equilibrium point.
The resulting $x$ curve tends to an exponentially damped sinusoid.
A couple of such trajectories in the phase space are plotted in Fig.~\ref{fig:ncspi}.
A stable point like that of the former type is called a {\it sink}, while a point of the
latter type is a {\it spiral} \citep[see][]{Strogatz1994}.
In this figure, equilibrium points are close to the minimum of the cubic nullcline, 
whose coordinates are $x_m = 2\beta/3 =2.0$ and $y_m = -4\beta^3/9 = -12.0$.
There is, therefore, a rather narrow interval $[x_2, x_m]$ where the equilibrium is stable 
and the system describes a relatively high number of converging rounds. 
In Fig.~\ref{fig:tr} this interval is limited by the second turquoise and the violet 
vertical lines; the corresponding interval for $J_0$ is $S_{2*} \equiv [5.99382..., 8.0]$ 
and it is reported as the QPO range, although it is possible to have this feature in the 
adjacent intervals.

In this paper, we are interested only to equilibrium points of spiral type which can
be related to the appearance of LFQPOs.
The typical decay time decreases very rapidly for $J_0$ increasing from the stability
limit to values slightly higher than $x_m$, thus when this parameter is between 6 and 7.5
the solution can have a rather long series of oscillations.
For $J_0 > 8.0$ the phase space trajectory has a small number of cycles and  for 
higher enough values ($\gtrsim 15.0$) the path does not encircle at all the equilibrium 
point. 
Note that also for such high $J_0$ value the corresponding $x_* = 2.4662..$ is quite close
to $x_m$.

It is easy to calculate a linear approximation of the MHR system of equation~\ref{eq:ode}  
that gives a rather good solution in this neighbourhood:

\begin{equation}
\left\{
\begin{array}{lll}
\dot{x} &=& \big(\frac{\partial f}{\partial x}\big)_m (x - x_m) + \big(\frac{\partial f}{\partial y}\big)_m (y - y_m) = y - y_m \\ 
  &  &  \\
\dot{y} &=& \big(\frac{\partial g}{\partial x}\big)_m (x - x_m) + \big(\frac{\partial g}{\partial y}\big)_m (y - y_m) = \\
  &  &  \\
        &=& -\frac{4}{3} \beta^2 (x - x_m) - (y - y_m)   %\nonumber
\end{array}\right.
\label{eq:partial}
\end{equation}

\noindent
where the partial derivative are evaluated at ($x_m, y_m$).
This linearized system is that of a damped harmonic oscillator:

\begin{equation}
 \ddot{w} + \dot{w} + \frac{4}{3} \beta^2 w = 0
\label{eq:armonic}
\end{equation}

\noindent
with $w = x -x_m$.
Thus the solution of the MHR model, when approaching the equilibrium point, can 
be well approximated by a sinusoid with an exponentially decreasing amplitude.
This is seen in Fig.~\ref{fig:ncspi} where the trajectory for $J_0 = 6.5$ evolves to 
an elliptical shape converging at the equilibrium.
It is interesting that the frequency of this oscillation is depending only on $\beta$ 
and results:

\begin{equation}
 \nu_0 = \frac{1}{2 \pi} \sqrt{4 \beta^2/3 -1/4}    ~~~~~
\label{eq:nu}
\end{equation}

\noindent
and the exponential decay time is equal to unity.
For $\beta > 2.0$ a very good approximation (better than 2\%) of this equation is 
$\nu_0 \approx  \beta /\pi \sqrt{3}$.

\begin{figure}
      \hspace{0cm}
%\epsfysize=12.0cm
%\epsfbox{A8phs.ps}
\includegraphics[height=7.5cm,angle=-90,scale=1.1]{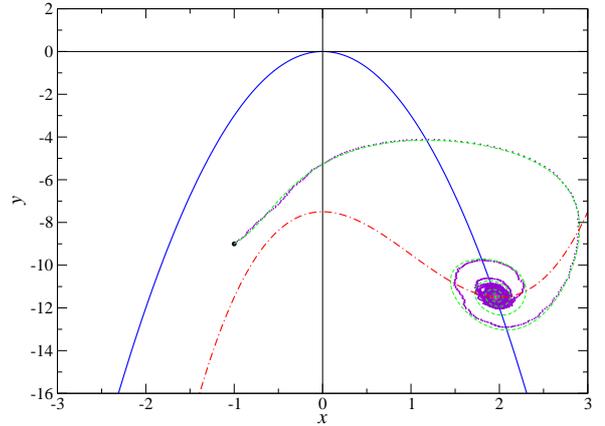}
\caption[]{
Numerical phase space trajectories of the MHR system with $\beta = 3$, $J_0 = 7.5$ and
with a random component of amplitude $C = 6.0$ (violet dotted curve) and without the 
random input (green dashed curve).
Red dash-dotted and solid blue curves are the cubic and the parabolic nullclines, 
respectively.
}
\label{fig:ntrspi}
\end{figure}

\begin{figure}
      \hspace{-0.5cm}
%\epsfysize=12.0cm
%\epsfbox{A8phs.ps}
\includegraphics[height=8.0cm,angle=-90,scale=1.1]{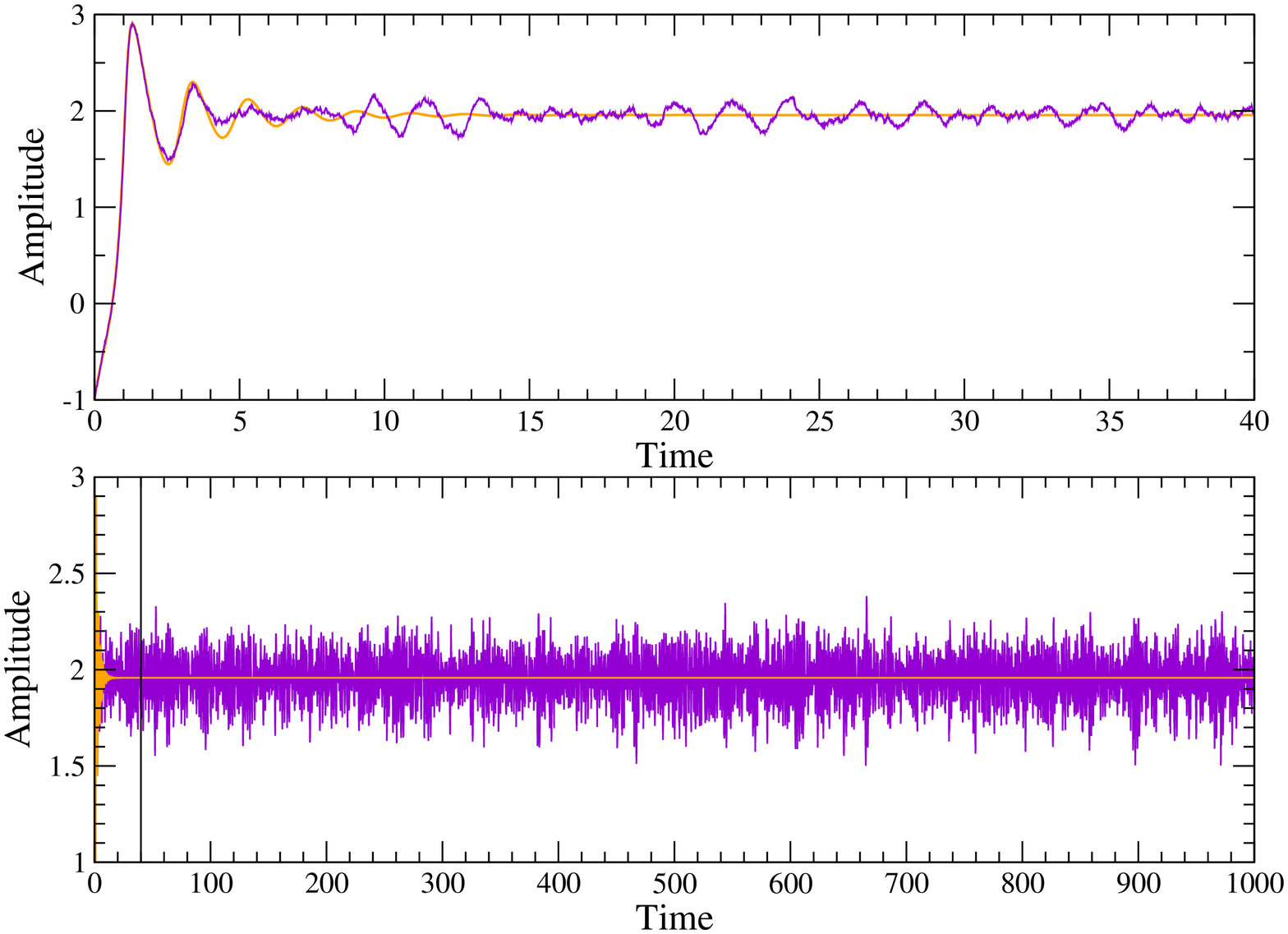} \\
\includegraphics[height=8.0cm,angle=-90,scale=1.1]{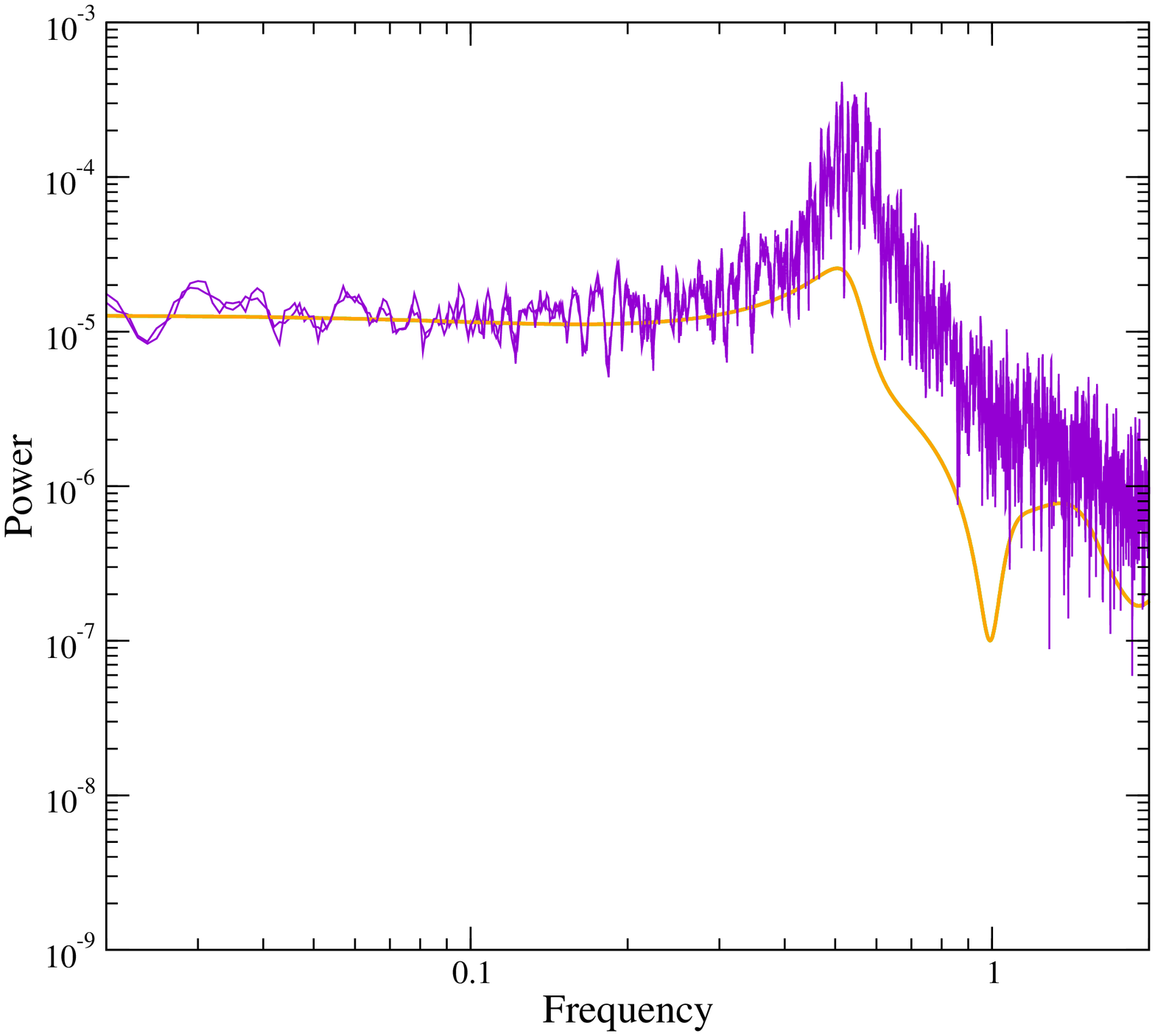}
\caption[]{
Upper plot: in the upper panel there is the initial segment of the numerical time solutions 
for the $x$ variable of the system of equation~\ref{eq:ode} with $\beta = 3$, $J_0 = 7.5$ 
and a random component amplitude $C = 6.0$ (violet dotted curve) and without the random 
input (orange thick curve); in the lower panel there is a longer light curve with 
the same parameters' values to show that the pattern remains stable, the vertical line
limits in the interval shown in the upper plot. 
Lower plot: Fourier PDS of the quasi-periodic oscillation after a 3 point running average,
the lower orange curve is the PDS of the signal without noise shown in in the same figure.
Note that the peak frequencies are the same, but the height of the latter spectrum is
much lower than the one of QPO signal.
}
\label{fig:qpolc}
\end{figure}

\subsection{Solutions with a random noise component}    
%%%%%%%%%%%%%%%%%%%    <<<<  SUBSECTION

We now study the solutions when a random noise component is added to the input function

\begin{equation}
 J(t) = J_0 + C r(t)  
 \label{eq:jfunc}
\end{equation}

\noindent
where $r(t)$ is a random number with a uniform distribution in the interval [$-$0.5, 0.5].
This term is present in the equation for $\dot{x}$ and therefore it implies that the cubic
nullcline in no longer stable but it is rapidly translating along the vertical axis around
its mean position that is the curve corresponding to the cubic with $J(t) =J_0$. 

The trajectory shown in Fig.~\ref{eq:jfunc} was computed for $J_0 = 7.5$ and $C = 6.0$: 
note that in this case the random term is large enough to move the equilibrium point into 
the unstable region.
The evolution of this trajectory, however, presents a transient in which the noise 
acts only as a small perturbation with respect to path in absence of noise.
However, when the trajectory approaches the mean equilibrium state, it does not converges
directly to this point but describes small approximately elliptical curves in 
its surroundings having a variable amplitude.
It appears as a `steady' situation because this oscillation continues definitively 
and has the aspect of a periodic signal with an amplitude modulated on a timescale 
ranging from about five to ten fundamental periods.
The effect of large random changes of $J(t)$ appears like a weak perturbation 
of the trajectory because these changes act only on the derivative of $x$, and the 
corresponding variations of this variable are too small to produce a large deviation
from the undisturbed path.
As a consequence, the cumulative effect of these fast changes is negligible and the
resulting trajectory exhibits only small deviations with respect to that corresponding
to the constant $J_0$.

The time scale of calculated signals was chosen to have QPO frequencies close 1 Hz.
The time evolution of this solution is given in the two  panels of the upper plot of 
Fig.~\ref{fig:qpolc}, where there is a detail of the lower plot to show the transient 
phase and the first segment of oscillating pattern.
We also reported the solution without the random noise to make clear the different 
behavior of the resulting curves when this component is considered.
The lower plot reports the corresponding PDS, where a LFQPO feature, remarkably 
similar to the one observed in \grss (see Fig.~\ref{fig:rxte_qpo_pds}), confirming t
hat in these conditions the MHR model can originate this phenomenon.
The resulting light curve is like those shown in the upper plot of Fig.~\ref{fig:cpsl} 
and therefore it is also remarkably similar to that derived from the filtered data.
This similarity reduces the PDS degeneracy, i.e. the fact that different types of 
signal have analogous PDS, and confirms that the MHR model reproduces the light curves
of these unstable states with a high accuracy.

\section{Unstable solutions, limit cycle and high frequency oscillations} 
%%%%%%%%%%    <<<<  SECTION  5

For values of of $J_0$ within the interval $[0.0061792,0.599382]$ the corresponding 
values of the equilibrium point are in the unstable interval and, as shown in 
\citetalias{Massaro2020a}, the MHR model describes a limit cycle whose light curve is 
like that of the $\rho$ class (see Fig.~\ref{fig:mhr_1}).
The period of the limit cycle varies regularly with $J_0$ according to a power law of
exponent 0.5: this variation is due to the shortening of the slow leading trail making
the signal shape more and more similar to a sinusoid.
In \citetalias{Massaro2020a} we named this particular pattern `high frequency oscillation`
(shortly `hfo') because its frequency corresponds to the highest one that one can reach 
increasing $J_0 \in I$ and that is slightly lower than the value estimated by means of 
equation~\ref{eq:nu}.

\begin{figure}
      \hspace{0cm}
%\epsfysize=12.0cm
%\epsfbox{A8phs.ps}
\includegraphics[height=8.0cm,angle=-90,scale=1.1]{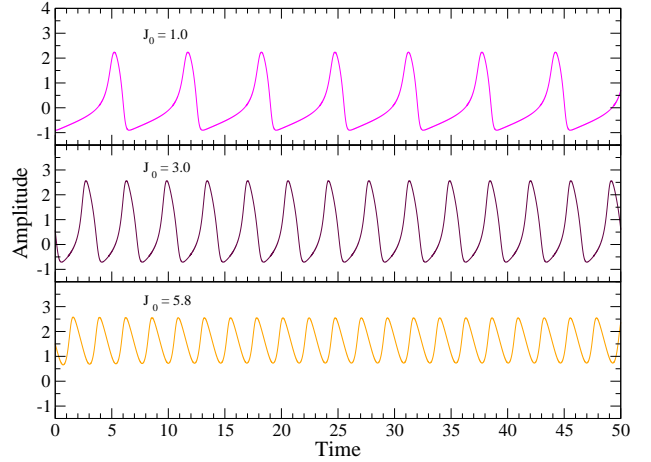}
\caption[]{
Three light curves computed with the MHR model with $\beta = 3.0$ for different values 
of $J_0$ to show the decrease of the period and the evolution of the shape towards an 
approximate sinusoid.
}
\label{fig:qpopds_a}
\end{figure}

\begin{figure}
      \hspace{0cm}
%\epsfysize=12.0cm
%\epsfbox{A8phs.ps}
\includegraphics[height=8.0cm,angle=-90,scale=1.1]{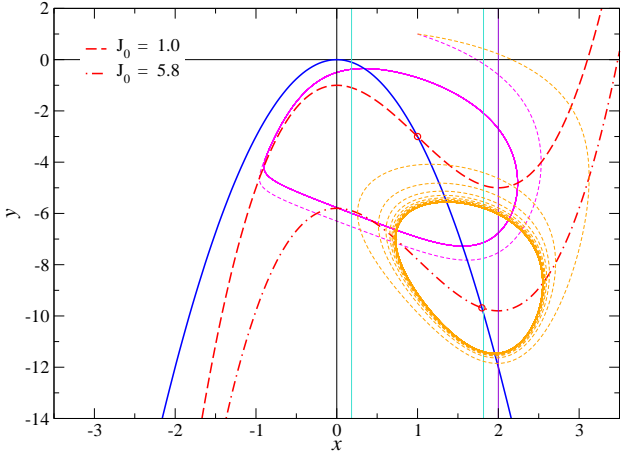}
\caption[]{
Phase space plot to show the nullclines (red and blue curves) and the trajectories 
(dashed lines) of two curves in Fig.~\ref{fig:qpopds_a}.
Turquoise vertical lines delimit the unstable interval and the violet line indicates
the coordinate of the cubic nullcline minimum.
}
\label{fig:qpopds_b}
\end{figure}

These two effect are clearly visible in the three curves of limit cycles reported in
Fig.~\ref{fig:qpopds_a}: note, in particular, the curve for $J_0 = 5.8$, a value close 
the upper boundary of the unstable interval, whose profile is approximating a sinusoidal 
shape.
The corresponding phase space trajectories are shown in Fig.~\ref{fig:qpopds_b} (only 
two trajectories are plotted to avoid confusion): both curves have sections in stable 
intervals and particularly the one with the higher $J_0$ is for about half cycle in 
the stable region.
Note also that the equilibrium point is located very close to the minimum of the cubic
nullcline and this confirms that equation~\ref{eq:nu} can be assumed as a valuable 
approximation for the `hfo' frequency.

\begin{figure}
      \hspace{0cm}
%\epsfysize=12.0cm
%\epsfbox{A8phs.ps}
\includegraphics[height=8.0cm,angle=-90,scale=1.1]{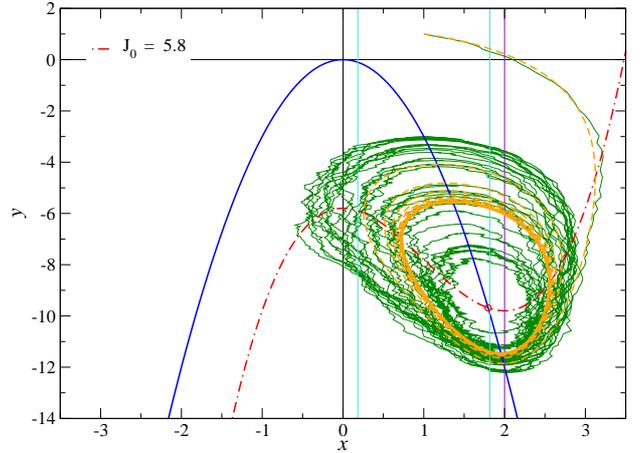}
\caption[]{
Phase space plot to show the nullclines (red dot-dashed and solid blue curves) 
and the trajectories of an MHR result for $J_0 = 5.8$ without (orange dashed line) 
and with (dark green line) a noise component with $C = 20.$. 
Turquoise vertical lines delimit the unstable interval and the violet line indicates
the minimum coordinate of the cubic nullcline.
}
\label{fig:qpopds_c}
\end{figure}

\begin{figure}
      \hspace{0cm}
%\epsfysize=12.0cm
%\epsfbox{A8phs.ps}
\includegraphics[height=8.0cm,angle=-90,scale=1.1]{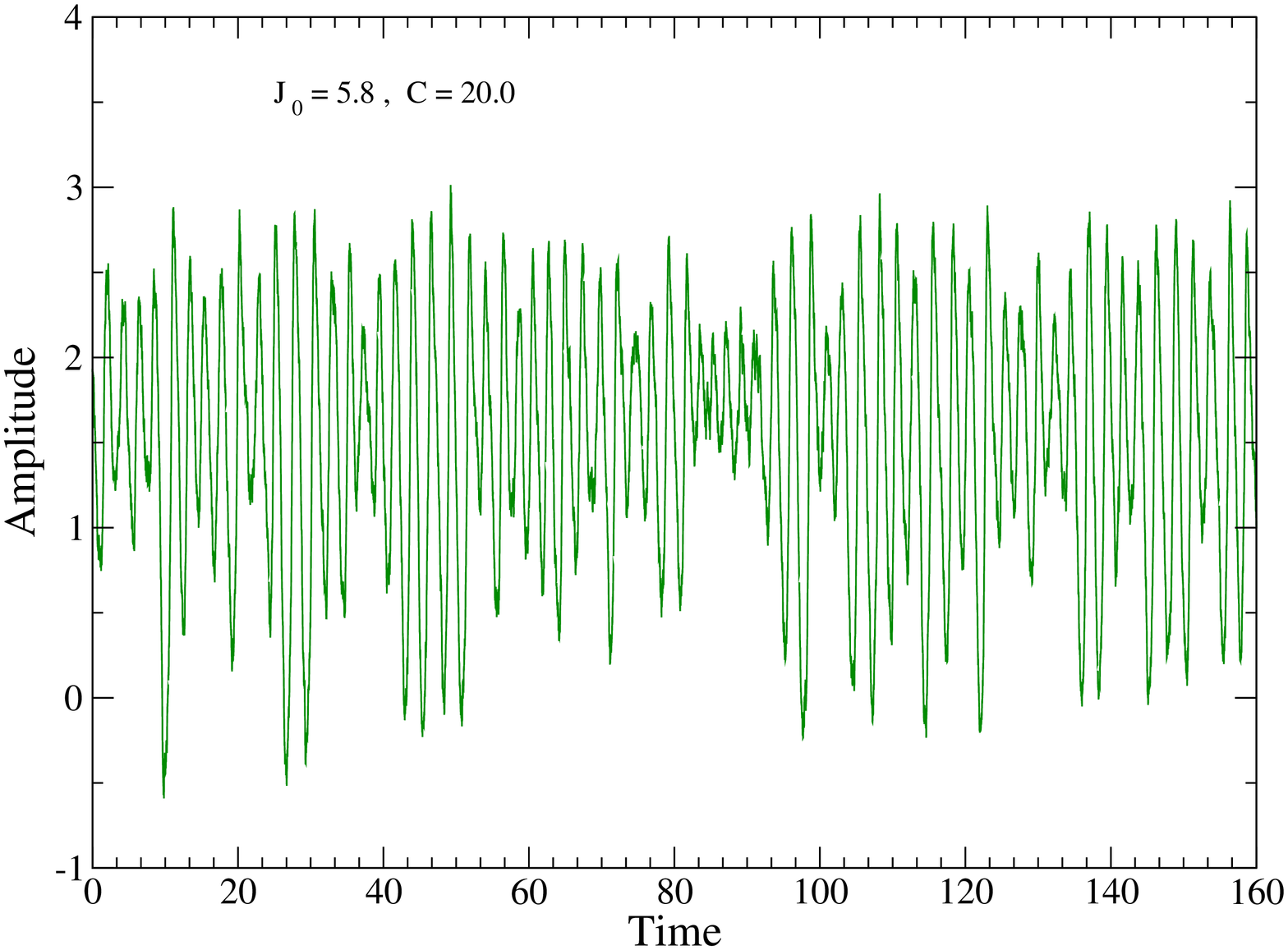} \\
\includegraphics[height=8.0cm,angle=-90,scale=1.1]{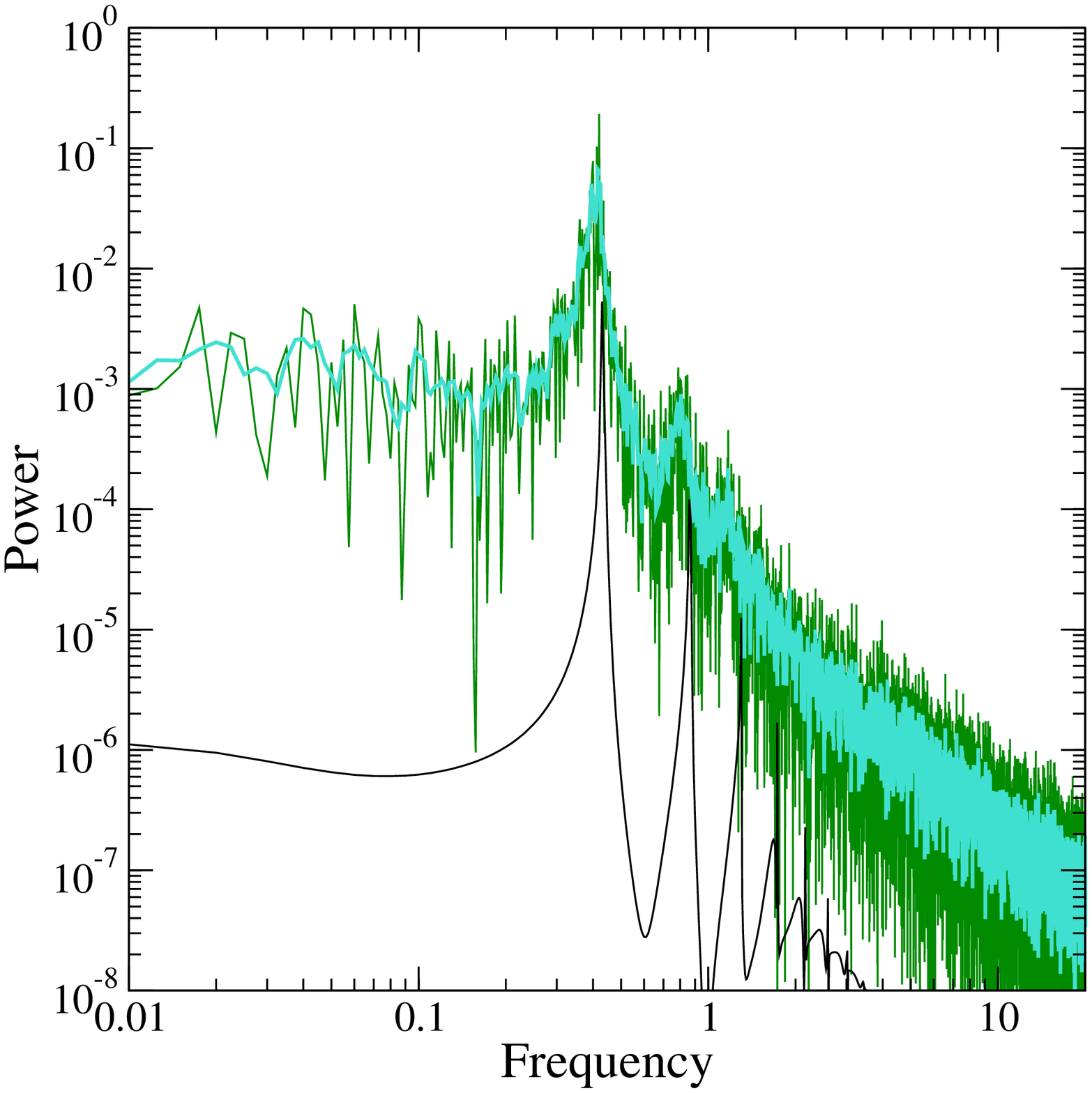} \\
\caption[]{
Upper plot: Light curve computed with the MHR model with $\beta = 3.0$ and $J_0 = 5.8$ 
and a large noise component ($C = 20$). 
Note the asymmetry of the signal with the noise fluctuations greater in the lower part 
of the curve than in the upper one. 
Lower plot: Fourier PDS of the signal in the upper panel (dark green data) showing a 
well evident broad peak and three harmonics; these features are more apparent after
a running average over 5 points (turquoise data).
The black spectrum is that of the signal without noise.
}
\label{fig:qpopds_d}
\end{figure}

As seen above, the addition of a low amplitude noise introduces only small perturbations 
in the resulting signals, but when this amplitude increases up to a value $C = 15$ or even 
higher the phase space trajectory (see Fig.~\ref{fig:qpopds_c}) exhibits a more complex 
pattern with large separated annular patterns, implying a low frequency modulation.
A short segment of the light curve is in the upper plot in Fig.~\ref{fig:qpopds_d},
where the amplitude modulation is evident.
In the lower plot of the same figure we report the PDS of the this signal exibiting a 
prominent broad peak again very similar to the LFQPO feature in Fig.~\ref{fig:rxte_qpo_pds}.
The solid black line is PDS of the `hfo' without noise.
The main peak and its harmonics are more evident after a light smoothing (turquoise 
data in the same figure) and its central frequency is slightly lower than that of 
`hfo' without noise.
As a further remark we underline that the structure of the noisy phase space 
trajectory in Fig.~\ref{fig:qpopds_c} is recalling the solutions of the Lorenz model
 \citep{Lorenz1963} and therefore it suggests that the chaotic behavior found in some 
light curves of \grss \citep{Misra2006} can be related to the same processes described by
the MHR model, that is not properly a chaotic deterministic system, but a noise perturbed 
limit cycle.

\section{Noise level and LFQPO intensity} 
%%%%%%%%%%%%%%%%%%%%%%%%%%%%%%    <<<<  SECTION  6
\label{sect:strqpo} 

Our results indicate that the occurrence of LFQPOs is dependent upon the presence 
of a noisy component: in other conditions, in fact, solutions exhibit a `hfo' or 
converge spiralling toward the equilibrium point according to the value of $J_0$, 
lower or higher the threshold stability, respectively.
We verified this connection performing some numerical calculations with different
choices of parameters and, in the following, we show the results of nine cases for 
three choices of $J_0$ and three of $C$. 
The former parameter was taken equal to 5.7 in the $I$ `hfo' interval, to 6.5, that 
is in $S_{2*}$, and to 8.0, so that the equilibrium point coincides with the local 
minimum of the cubic nullcline.
The three values of $C$ are 1.0, 5.0 and 9.0.

\begin{figure}\centering
\includegraphics[height=9.3cm,angle=-90,scale=1.0]{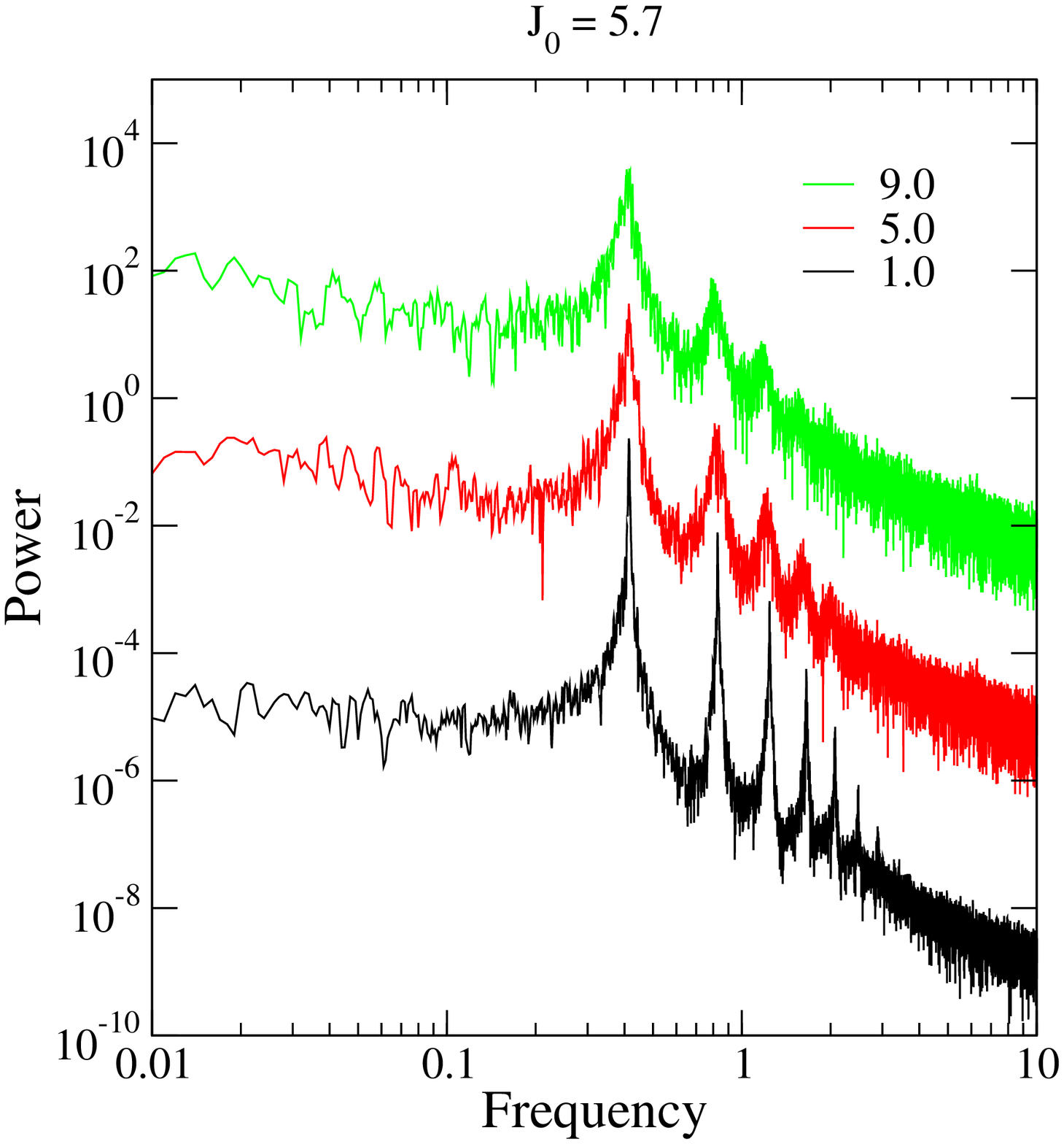}  \\
\includegraphics[height=9.3cm,angle=-90,scale=1.0]{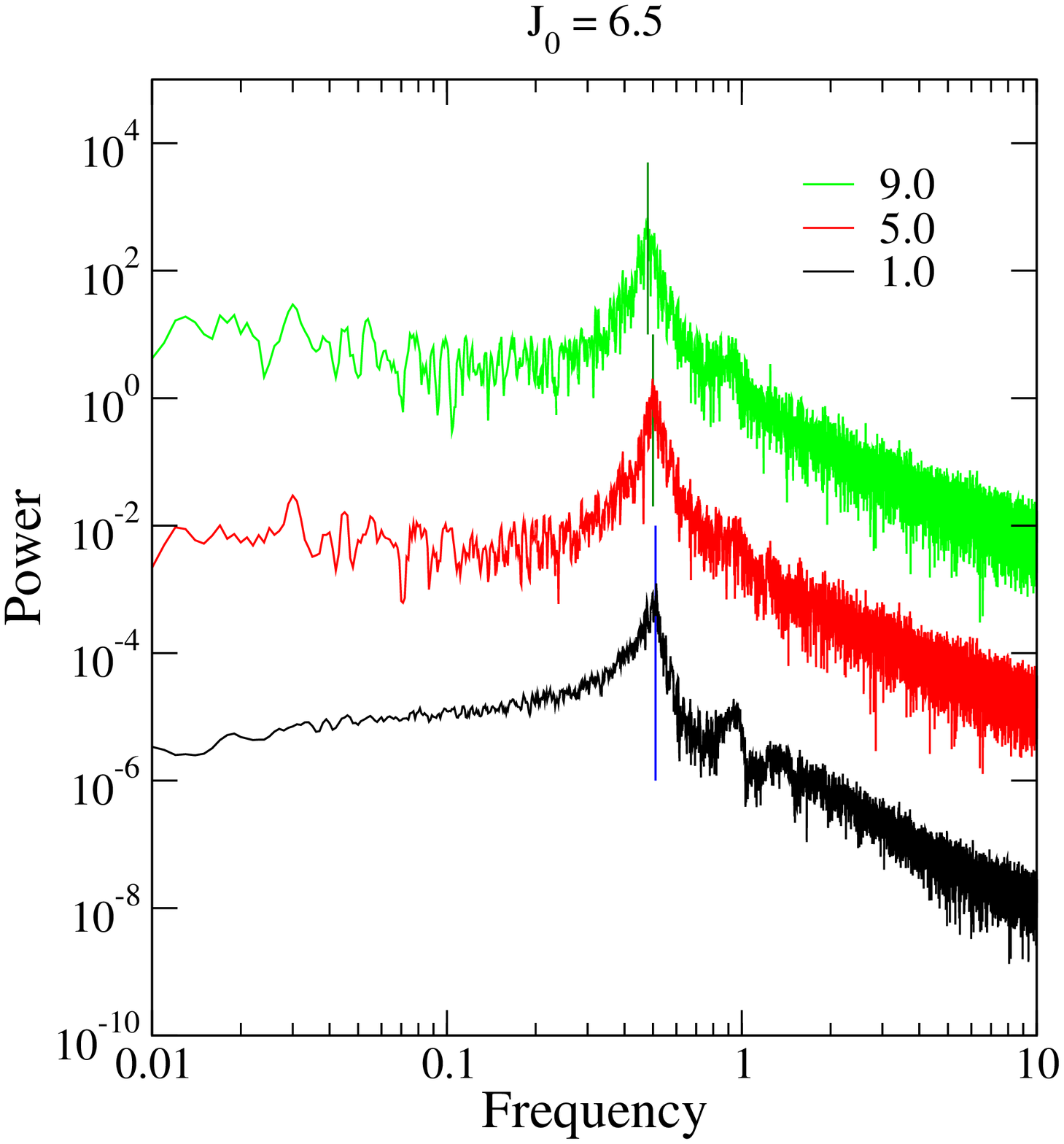} 
\includegraphics[height=9.3cm,angle=-90,scale=1.0]{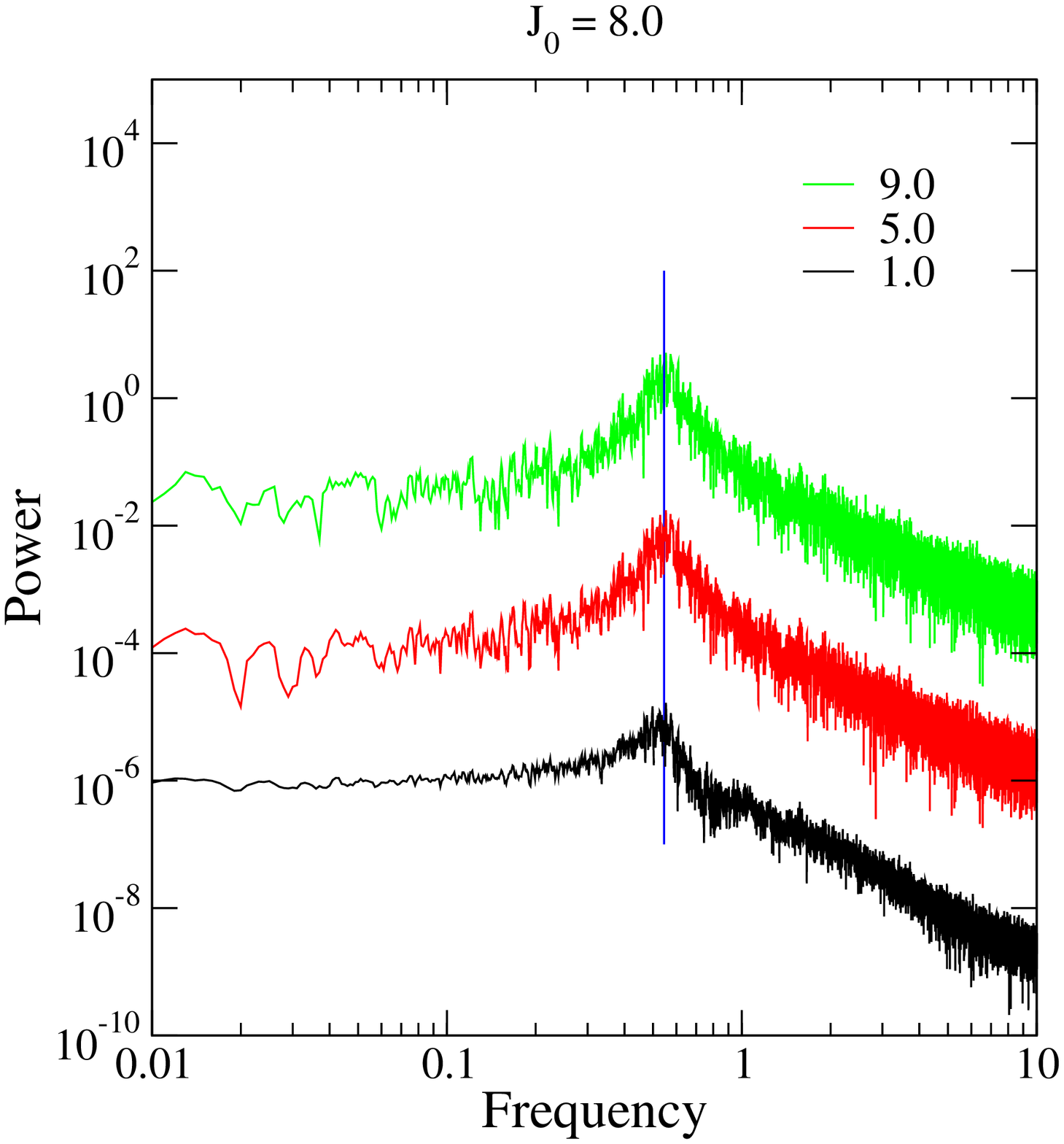} 
\caption[]{
Top plot: PDS computed by means of MHR model for $\beta = 3.0$ and $J_0 = 5.7$
and three different noise levels; spectra were vertically shifted to avoid confusion. \\
Central plot: PDS computed for $J_0 = 6.5$; vertical lines mark the central frequencies 
of LFQPO peaks. \\
Bottom plot: PDS computed for $J_0 = 8.0$.}
\label{fig:qpostruct}
\end{figure}

The resulting PDS are given in the three panels in Fig.~\ref{fig:qpostruct}. 
The top panel shows the PDSs for the `hfo' state with increasing added noise: 
the spectrum of the lowest noise data exhibits a harmonic series of narrow peaks 
about two orders of magnitude higher than the noise amplitude.
A lower and lower number of harmonics appears also when the noise increases but 
the central frequency and width remain stable to the `hfo' value, that is equal
to 0.41 in the units corresponding to those adopted for the time.
For $J_0$ high enough to move the system in the stable region and a QPO feature 
is always present in the PDS, but with only one or two harmonics.
Its central frequency decreases slightly for increasing noise from 0.51 to 0.48, 
in any case higher by about 20\% than in the previous case.
A further increase of $J_0$ moves the equilibrium point at the cubic local 
minimum and the PDS continues to present the QPO peak but it appears mild and 
with the only the first harmonic barely detectable; its central frequency is 
stable at 0.545, as expected from equation~\ref{eq:nu}.

\section{The origin of LFQPOs in the MHR model}
 %%%%%%%%%%%%%%%%%%%%%%%%%%%%%%      <<<<  SECTION  7
\label{sect:orqpo} 

As shown in the introduction, the MHR model reproduces well light curves of several 
variability classes according to the values of the single parameter $J_0$, which 
controls transitions from stable to unstable equilibrium.
In the latter states the system describes a limit cycle whose period decreases
for increasing $J_0$ and the spike profiles change to a more and more symmetric
shape that approximates a sinusoid with a short and constant period, here
named `high frequency oscillations'.
A further increase of $J_0$ produces a transition to the second stable region 
thus we expect a signal evolving to a steady level.
In Sect.~\ref{sct:stable_solut} we discussed the nature of the equilibrium points 
and showed that they are of type {\it sink} of of type {\it spiral} according that 
the value of $J_0$ is in $S_1$ or $S_2$, respectively.
In the latter case the $x$ curve presents a number of oscillations before to 
reach the final value.
Fast fluctuations of the $J(t)$ generally act on the phase space trajectories as 
small perturbations with respect to the one corresponding to the mean value $J_0$.
For values close to the upper boundary of the unstable interval $I$, the system
excite `hfo' modes with an amplitude modulation on longer time scales.
The corresponding PDS shows a broad feature, typical of LFQPOs frequently found 
in BHCs, like \grs.
When $J_0$ values are above the stability threshold the equilibrium point remains 
always in the stable region, and LFQPOs are also present if the $x$ coordinate of
this point is lower than or close to the minimum of the cubic nullcline. 
These results allow us to favor an  ''intrinsic'' hypothesis on the origin of 
LFQPOs in an accretion disc essentially related to the same mechanism responsible 
of the spiking limit cycle and occurring for $J_0$ values close to transition 
between the unstable and the stable region.

The role of random fluctuations, that could be due to plasma turbulence in the
disc, in establishing LFQPOs was already noticed in a paper by
\citet{Maccarone2011}, who computed the bispectrum of some observations of 
\grss and found a correlation between LFQPOs and the variations of the noise 
component.
This tool provides the possibility of discriminating among the various variability 
modes producing similar PDS and \citet{Maccarone2011} concluded that ``the 
variability is caused by a reservoir of energy being drained by a noise component 
... and a quasi-periodical component, while in the brighter part of the $\chi$ state, 
the variability is consistent with a white noise input spectrum driving a damped 
harmonic oscillator with a non-linear restoring force.''
MHR results are in agreement with this finding and confirm the relevance of the
noise, in particular we found that small deviations from the decaying trajectory 
perturb it towards a different path which converges again to the equilibrium until 
another small deviation restores a similar condition.
Then the noise is like a stabilizing factor for the LFQPO and their frequency 
is limited in a rather narrow interval close to the one of the corresponding 
oscillator at the local minimum of the cubic as shown in Sect.~\ref{sct:stable_solut}.

The correlations of LFQPO frequency with the photon energy of the source luminosity
are useful for addressing possible relationships with the MHR model parameters.
For example, according equation~\ref{eq:nu}, these correlations would imply that 
the parameter $\beta$ must be linear depending on the photon energy.

It is important to point out that our results does not exclude geometric models for 
LFQPOs, particularly in some sources which do not exhibit the same complex variability 
of \grs.
These models can naturally account for some phenomena as the modulation of the iron 
line energy as a function of QPO phase \citep[see, for instance,][]{Ingram2016,Nathan2019}.  
The present version of the MHR model is only focused on the time stucture of the brightness 
changes and does not include any energy dependence of the emission and, particularly, the
properties of the iron line.
We stress that it should be considered as a simple tool approximating the non-linear 
instability in accretion discs which produces the large variety of light curves, as
those observed in \grs.
It may have, however, a heuristic content because can help the understanding of some
features and details, like the spike profiles or the LFQPO signal structure.

\section{Conclusion}  
%%%%%%%%%%%%%%%%%%%%%%%%%%%%%%%%%%%%%%%%%%%%%%%%%%  <<<<<<<  SECTION  8
In \citetalias{Massaro2020a, Massaro2020b} we proposed the non linear mathematical 
MHR model, containing only a small number of parameters, whose solutions reproduce 
several different classes of light curves of \grss, and describe well the 
transition from stables to bursting states.
An interesting finding of this model was that it is also able to describe the
occurrence of LFQPOs as a consequence of a transition from an unstable to
a stable equilibrium.
In the present paper we studied in detail the nature of this transition and
compare the model results with some observational data.

The major findings of the present work are related to the fact that the stable
equilibrium points where LFQPOs are present is of the {\it spiral} type.
Moreover, we found that for values of the driving parameter $J_0$ within the 
interval $S_{2*}$, the equilibrium point lies between the stability threshold 
and the local minimum of the cubic nullcline.
In this condition the phase space trajectories converge to the equilibrium point
describing a tight spiral around it, that corresponds to an oscillating
pattern in the model light curve.
It follows that the PSDs are very similar to the observed ones with a broad 
Lorentzian feature and, occasionally, one or two harmonics.
The general structure of model light curves is also like the observed ones after
a filtering in the peak range.

Another important finding is that the fluctuations of $J_0$ play a role in 
stabilizing LFQPOs: without noise the phase space trajectory converges to the 
equilibrium and light curves are like a damped oscillation, while random 
displacements can move the trajectory towards outer positions from which 
a new path approaching to equilibrium follows.
Without noise the occurrence of long duration LFQPOs would not be possible.
This result confirms the bispectral analysis of some light curves of \grss by
\citet{Maccarone2011} who pointed out the noise relevance in the process 
responsible of LFQPOs.
The possibility of noise-induced quasi-periodic oscillation in the original 
HR model including three ODEs was also considered by \citet{Ryashko2017},
confirming thus the relevance of the random fluctuations although in different 
conditions.
Turbulence in accretion discs is important because it can provide a driving
mechanism also for HFQPOs and can produce the 3:2 twin peak feature, as  
demonstrated by the numerical model recently developed by \citet{Ortega2020}.
In Sect.~\ref{sct:ode} we derived a simple linear approximation of the MHR 
model that was used for estimating the central frequency of the Lorentzian 
peak in a narrow interval around the local minimum of the cubic nullcline 
that was found to be depending only on the parameter $\beta$.
More generally, one could expect that this frequency is determined by the 
shape of equilibrium track near the minimum and, if this curve can be derived 
from a physical stability calculations as made, for instance by \citet{Watarai2001}
\citepalias[see][]{Massaro2020b}, one could relate the observed 
LFQPO data to some parameters of the accretion disc.

These results suggest a possible explanation why many BHCs exhibit LFQPOs but
not the large variety of light curve profiles as those of \grs.
In fact, it would be sufficient that the values of the equivalent $J_0$ in 
these sources remain for all the time in the range corresponding to an
equilibrium point close or just above the boundary between the unstable 
interval of `hfo' and spiral trajectories.
An interesting property worthwhile of investigation is if such a condition 
can be related to their disc sizes which for many BHCs are estimated much 
lower than \grss \citep{Remillard2006}.

The present results confirm that MHR model is a simple and efficient 
approximation for describing the instabilities in an accretion disc and 
predicting a large variety of light curves that are originated in this 
type of physical processes. 
Using this model we also showed that the origin of LFQPOs can be explained 
by the same instability, but for values of the input parameter in a range
higher than the unstable interval. 
There are, however, some relevant topics to further study:
the most important is to complete the physical interpretation of the model
and the association of the mathematical variables with physical quantities
of the accretion disc.
This association requires numerical calculations of equilibrium states and
an analysis of their stability using hydrodynamic codes.
It is also possible that in this way one will open the possibility of adding 
the variable energy and of achieving a more complete mathematical modelling
of the source behaviour in different bands.

\section*{Acknowledgments}
The authors are grateful to Marco Salvati and Andrea Tramacere for their 
fruitful comments.
MF, TM and FC acknowledge financial contribution from the agreement ASI-INAF 
n.2017-14-H.0
%%%%%%%%%%%%%%%%%%%%%%%%%%%%%%%%%%%%%%%%%%%%%%%%%%

%%%%%%%%%%%%%%%%%%%% REFERENCES %%%%%%%%%%%%%%%%%%
\section*{Data availability}
Data used in this paper are available in a repository and can be accessed via link https://heasarc.gsfc.nasa.gov/cgi-bin/W3Browse/w3browse.pl.

\bibliographystyle{mnras}
\bibliography{grs1915_III} 

\begin{thebibliography}{}
\makeatletter
\relax
\def\mn@urlcharsother{\let\do\@makeother \do\$\do\&\do\#\do\^\do\_\do\%\do\~}
\def\mn@doi{\begingroup\mn@urlcharsother \@ifnextchar [ {\mn@doi@}
  {\mn@doi@[]}}
\def\mn@doi@[#1]#2{\def\@tempa{#1}\ifx\@tempa\@empty \href
  {http://dx.doi.org/#2} {doi:#2}\else \href {http://dx.doi.org/#2} {#1}\fi
  \endgroup}
\def\mn@eprint#1#2{\mn@eprint@#1:#2::\@nil}
\def\mn@eprint@arXiv#1{\href {http://arxiv.org/abs/#1} {{\tt arXiv:#1}}}
\def\mn@eprint@dblp#1{\href {http://dblp.uni-trier.de/rec/bibtex/#1.xml}
  {dblp:#1}}
\def\mn@eprint@#1:#2:#3:#4\@nil{\def\@tempa {#1}\def\@tempb {#2}\def\@tempc
  {#3}\ifx \@tempc \@empty \let \@tempc \@tempb \let \@tempb \@tempa \fi \ifx
  \@tempb \@empty \def\@tempb {arXiv}\fi \@ifundefined
  {mn@eprint@\@tempb}{\@tempb:\@tempc}{\expandafter \expandafter \csname
  mn@eprint@\@tempb\endcsname \expandafter{\@tempc}}}

\bibitem[\protect\citeauthoryear{{Belloni}, {Klein-Wolt}, {M{\'e}ndez}, {van
  der Klis}  \& {van Paradijs}}{{Belloni} et~al.}{2000}]{Belloni2000}
{Belloni} T.,  {Klein-Wolt} M.,  {M{\'e}ndez} M.,  {van der Klis} M.,   {van
  Paradijs} J.,  2000, A\&A, \href
  {http://adsabs.harvard.edu/abs/2000A\%26A...355..271B} {355, 271}

\bibitem[\protect\citeauthoryear{{Casella}, {Belloni}, {Homan}  \&
  {Stella}}{{Casella} et~al.}{2004}]{Casella2004}
{Casella} P.,  {Belloni} T.,  {Homan} J.,   {Stella} L.,  2004, \mn@doi [\aap]
  {10.1051/0004-6361:20041231}, \href
  {https://ui.adsabs.harvard.edu/abs/2004A&A...426..587C} {426, 587}

\bibitem[\protect\citeauthoryear{{Casella}, {Belloni}  \& {Stella}}{{Casella}
  et~al.}{2005}]{Casella2005}
{Casella} P.,  {Belloni} T.,   {Stella} L.,  2005, \mn@doi [\apj]
  {10.1086/431174}, \href
  {https://ui.adsabs.harvard.edu/abs/2005ApJ...629..403C} {629, 403}

\bibitem[\protect\citeauthoryear{{Castro-Tirado}, {Brandt}  \&
  {Lund}}{{Castro-Tirado} et~al.}{1992}]{CastroTirado1992}
{Castro-Tirado} A.~J.,  {Brandt} S.,   {Lund} N.,  1992, IAUC, \href
  {http://adsabs.harvard.edu/abs/1992IAUC.5590....2C} {5590, 2}

\bibitem[\protect\citeauthoryear{{Chakrabarti} \& {Molteni}}{{Chakrabarti} \&
  {Molteni}}{1993}]{Chakrabarti1993}
{Chakrabarti} S.~K.,  {Molteni} D.,  1993, \mn@doi [\apj] {10.1086/173345},
  \href {https://ui.adsabs.harvard.edu/abs/1993ApJ...417..671C} {417, 671}

\bibitem[\protect\citeauthoryear{{Chen} \& {Taam}}{{Chen} \&
  {Taam}}{1992}]{Chen1992}
{Chen} X.,  {Taam} R.~E.,  1992, \mn@doi [\mnras] {10.1093/mnras/255.1.51},
  \href {https://ui.adsabs.harvard.edu/abs/1992MNRAS.255...51C} {255, 51}

\bibitem[\protect\citeauthoryear{{Chen} \& {Taam}}{{Chen} \&
  {Taam}}{1995}]{Chen1995}
{Chen} X.,  {Taam} R.~E.,  1995, \mn@doi [\apj] {10.1086/175360}, \href
  {https://ui.adsabs.harvard.edu/abs/1995ApJ...441..354C} {441, 354}

\bibitem[\protect\citeauthoryear{{Fender} \& {Belloni}}{{Fender} \&
  {Belloni}}{2004}]{Fender2004}
{Fender} R.,  {Belloni} T.,  2004, \mn@doi [\araa]
  {10.1146/annurev.astro.42.053102.134031}, \href
  {http://adsabs.harvard.edu/abs/2004ARA%26A..42..317F} {42, 317}

\bibitem[\protect\citeauthoryear{{Hindmarsh} \& {Cornelius}}{{Hindmarsh} \&
  {Cornelius}}{2005}]{Hindmarsh2005}
{Hindmarsh} J.~L.,  {Cornelius} P.,  2005, 2005, BURSTING: The Genesis of
  Rhythm in the Nervous System.
S. Coombes \& P.C. Bressloff eds

\bibitem[\protect\citeauthoryear{{Ingram} \& {van der Klis}}{{Ingram} \& {van
  der Klis}}{2015}]{Ingram2015}
{Ingram} A.,  {van der Klis} M.,  2015, \mn@doi [\mnras]
  {10.1093/mnras/stu2373}, \href
  {https://ui.adsabs.harvard.edu/abs/2015MNRAS.446.3516I} {446, 3516}

\bibitem[\protect\citeauthoryear{{Ingram}, {Done}  \& {Fragile}}{{Ingram}
  et~al.}{2009}]{Ingram2009}
{Ingram} A.,  {Done} C.,   {Fragile} P.~C.,  2009, \mn@doi [\mnras]
  {10.1111/j.1745-3933.2009.00693.x}, \href
  {https://ui.adsabs.harvard.edu/abs/2009MNRAS.397L.101I} {397, L101}

\bibitem[\protect\citeauthoryear{{Ingram}, {van der Klis}, {Middleton}, {Done},
  {Altamirano}, {Heil}, {Uttley}  \& {Axelsson}}{{Ingram}
  et~al.}{2016}]{Ingram2016}
{Ingram} A.,  {van der Klis} M.,  {Middleton} M.,  {Done} C.,  {Altamirano} D.,
   {Heil} L.,  {Uttley} P.,   {Axelsson} M.,  2016, \mn@doi [\mnras]
  {10.1093/mnras/stw1245}, \href
  {https://ui.adsabs.harvard.edu/abs/2016MNRAS.461.1967I} {461, 1967}

\bibitem[\protect\citeauthoryear{{Lasota}}{{Lasota}}{2016}]{Lasota2016}
{Lasota} J.,  2016, {Astrophysics of Black Holes - From fundamental aspects to
  latest developments}.
Springer-Verlag Berlin Heidelberg, \mn@doi{10.1007/978-3-662-52859-4}

\bibitem[\protect\citeauthoryear{{Lorenz}}{{Lorenz}}{1963}]{Lorenz1963}
{Lorenz} E.~N.,  1963, \mn@doi [Journal of Atmospheric Sciences]
  {10.1175/1520-0469(1963)020<0130:DNF>2.0.CO;2}, \href
  {https://ui.adsabs.harvard.edu/abs/1963JAtS...20..130L} {20, 130}

\bibitem[\protect\citeauthoryear{{Maccarone}, {Uttley}, {van der Klis},
  {Wijnands}  \& {Coppi}}{{Maccarone} et~al.}{2011}]{Maccarone2011}
{Maccarone} T.~J.,  {Uttley} P.,  {van der Klis} M.,  {Wijnands} R. A.~D.,
  {Coppi} P.~S.,  2011, \mn@doi [\mnras] {10.1111/j.1365-2966.2011.18273.x},
  \href {https://ui.adsabs.harvard.edu/abs/2011MNRAS.413.1819M} {413, 1819}

\bibitem[\protect\citeauthoryear{{Marcel} et~al.,}{{Marcel}
  et~al.}{2020}]{Marcel2020}
{Marcel} G.,  et~al., 2020, arXiv e-prints, \href
  {https://ui.adsabs.harvard.edu/abs/2020arXiv200510359M} {p. arXiv:2005.10359}

\bibitem[\protect\citeauthoryear{{Markwardt}, {Swank}  \& {Taam}}{{Markwardt}
  et~al.}{1999}]{Markwardt1999}
{Markwardt} C.~B.,  {Swank} J.~H.,   {Taam} R.~E.,  1999, \mn@doi [\apjl]
  {10.1086/311899}, \href
  {https://ui.adsabs.harvard.edu/abs/1999ApJ...513L..37M} {513, L37}

\bibitem[\protect\citeauthoryear{{Massaro}, {Feroci}, {Mineo}, {Ardito}  \&
  {Ricciardi}}{{Massaro} et~al.}{2020a}]{Massaro2020a}
{Massaro} E.~{Capitanio} F.,  {Feroci} M.,  {Mineo} T.,  {Ardito} A.,
  {Ricciardi} P.,  2020a, \mnras, in press, (Paper I)

\bibitem[\protect\citeauthoryear{{Massaro}, {Feroci}, {Mineo}, {Ardito}  \&
  {Ricciardi}}{{Massaro} et~al.}{2020b}]{Massaro2020b}
{Massaro} E.~{Capitanio} F.,  {Feroci} M.,  {Mineo} T.,  {Ardito} A.,
  {Ricciardi} P.,  2020b, \mnras, in press, (Paper II)

\bibitem[\protect\citeauthoryear{{Misra}, {Harikrishnan}, {Ambika}  \&
  {Kembhavi}}{{Misra} et~al.}{2006}]{Misra2006}
{Misra} R.,  {Harikrishnan} K.~P.,  {Ambika} G.,   {Kembhavi} A.~K.,  2006,
  \mn@doi [\apj] {10.1086/503094}, \href
  {https://ui.adsabs.harvard.edu/abs/2006ApJ...643.1114M} {643, 1114}

\bibitem[\protect\citeauthoryear{{Morgan}, {Remillard}  \& {Greiner}}{{Morgan}
  et~al.}{1997}]{Morgan1997}
{Morgan} E.~H.,  {Remillard} R.~A.,   {Greiner} J.,  1997, \mn@doi [\apj]
  {10.1086/304191}, \href
  {https://ui.adsabs.harvard.edu/abs/1997ApJ...482..993M} {482, 993}

\bibitem[\protect\citeauthoryear{{Motta}}{{Motta}}{2016}]{Motta2016}
{Motta} S.~E.,  2016, \mn@doi [Astronomische Nachrichten]
  {10.1002/asna.201612320}, \href
  {https://ui.adsabs.harvard.edu/abs/2016AN....337..398M} {337, 398}

\bibitem[\protect\citeauthoryear{{Muno}, {Morgan}  \& {Remillard}}{{Muno}
  et~al.}{1999}]{Muno1999}
{Muno} M.~P.,  {Morgan} E.~H.,   {Remillard} R.~A.,  1999, \mn@doi [\apj]
  {10.1086/308063}, \href
  {https://ui.adsabs.harvard.edu/abs/1999ApJ...527..321M} {527, 321}

\bibitem[\protect\citeauthoryear{{Nathan}, {Ingram}, {Homan}  \&
  {Uttley}}{{Nathan} et~al.}{2019}]{Nathan2019}
{Nathan} E.,  {Ingram} A.,  {Homan} J.,   {Uttley} P.,  2019, Proc. X-ray
  Astronomy 2019 - Current Challenges and new frontiers in the Next Decade,
  Bologna 2019, Contr. 5500, \href
  {https://indico.ict.inaf.it/event/720/contributions/5500/} {}

\bibitem[\protect\citeauthoryear{{Ortega-Rodr{\'\i}guez},
  {Sol{\'\i}s-S{\'a}nchez}, {{\'A}lvarez-Garc{\'\i}a}  \&
  {Dodero-Rojas}}{{Ortega-Rodr{\'\i}guez} et~al.}{2020}]{Ortega2020}
{Ortega-Rodr{\'\i}guez} M.,  {Sol{\'\i}s-S{\'a}nchez} H.,
  {{\'A}lvarez-Garc{\'\i}a} L.,   {Dodero-Rojas} E.,  2020, \mn@doi [\mnras]
  {10.1093/mnras/stz3541}, \href
  {https://ui.adsabs.harvard.edu/abs/2020MNRAS.492.1755O} {492, 1755}

\bibitem[\protect\citeauthoryear{{Paul}, {Agrawal}, {Rao}, {Vahia}, {Yadav},
  {Marar}, {Seetha}  \& {Kasturirangan}}{{Paul} et~al.}{1997}]{Paul1997}
{Paul} B.,  {Agrawal} P.~C.,  {Rao} A.~R.,  {Vahia} M.~N.,  {Yadav} J.~S.,
  {Marar} T.~M.~K.,  {Seetha} S.,   {Kasturirangan} K.,  1997, \aap, \href
  {https://ui.adsabs.harvard.edu/abs/1997A&A...320L..37P} {320, L37}

\bibitem[\protect\citeauthoryear{{Press}, {Teukolsky}, {Vetterling}  \&
  {Flannery}}{{Press} et~al.}{2007}]{Press2007}
{Press} W.~H.,  {Teukolsky} S.~.,  {Vetterling} W.~T.,   {Flannery} B.~P.,
  2007, Numerical Recipes: The Art of Scientific Computing, 3 edn.
Cambridge University Press

\bibitem[\protect\citeauthoryear{{Remillard} \& {McClintock}}{{Remillard} \&
  {McClintock}}{2006}]{Remillard2006}
{Remillard} R.~A.,  {McClintock} J.~E.,  2006, \mn@doi [\araa]
  {10.1146/annurev.astro.44.051905.092532}, \href
  {https://ui.adsabs.harvard.edu/abs/2006ARA&A..44...49R} {44, 49}

\bibitem[\protect\citeauthoryear{{Rodriguez}, {Durouchoux}, {Mirabel}, {Ueda},
  {Tagger}  \& {Yamaoka}}{{Rodriguez} et~al.}{2002}]{Rodriguez2002}
{Rodriguez} J.,  {Durouchoux} P.,  {Mirabel} I.~F.,  {Ueda} Y.,  {Tagger} M.,
  {Yamaoka} K.,  2002, \mn@doi [\aap] {10.1051/0004-6361:20020218}, \href
  {https://ui.adsabs.harvard.edu/abs/2002A&A...386..271R} {386, 271}

\bibitem[\protect\citeauthoryear{{Ryashko} \& {Slepukhina}}{{Ryashko} \&
  {Slepukhina}}{2017}]{Ryashko2017}
{Ryashko} L.,  {Slepukhina} E.,  2017, \mn@doi [\pre]
  {10.1103/PhysRevE.96.032212}, \href
  {https://ui.adsabs.harvard.edu/abs/2017PhRvE..96c2212R} {96, 032212}

\bibitem[\protect\citeauthoryear{{Shilnikov} \& {Kolomiets}}{{Shilnikov} \&
  {Kolomiets}}{2008}]{Shilnikov2008}
{Shilnikov} A.,  {Kolomiets} M.,  2008, \mn@doi [International Journal of
  Bifurcation and Chaos] {10.1142/S0218127408021634}, \href
  {https://ui.adsabs.harvard.edu/abs/2008IJBC...18.2141S} {18, 2141}

\bibitem[\protect\citeauthoryear{{Strogatz}}{{Strogatz}}{1994}]{Strogatz1994}
{Strogatz} S.~H.,  1994, Nonlinear Dynamics and Chaos.
Westview Perseus Books Group, Reading MA

\bibitem[\protect\citeauthoryear{{Szuszkiewicz} \& {Miller}}{{Szuszkiewicz} \&
  {Miller}}{1998}]{Szuszkiewicz1998}
{Szuszkiewicz} E.,  {Miller} J.~C.,  1998, \mn@doi [\mnras]
  {10.1046/j.1365-8711.1998.01668.x}, \href
  {https://ui.adsabs.harvard.edu/abs/1998MNRAS.298..888S} {298, 888}

\bibitem[\protect\citeauthoryear{{Taam} \& {Lin}}{{Taam} \&
  {Lin}}{1984}]{Taam1984}
{Taam} R.~E.,  {Lin} D.~N.~C.,  1984, \mn@doi [\apj] {10.1086/162734}, \href
  {http://adsabs.harvard.edu/abs/1984ApJ...287..761T} {287, 761}

\bibitem[\protect\citeauthoryear{{Taam}, {Chen}  \& {Swank}}{{Taam}
  et~al.}{1997}]{Taam1997}
{Taam} R.~E.,  {Chen} X.,   {Swank} J.~H.,  1997, \mn@doi [\apjl]
  {10.1086/310812}, \href
  {https://ui.adsabs.harvard.edu/abs/1997ApJ...485L..83T} {485, L83}

\bibitem[\protect\citeauthoryear{{Tagger} \& {Pellat}}{{Tagger} \&
  {Pellat}}{1999}]{Tagger1999}
{Tagger} M.,  {Pellat} R.,  1999, \aap, \href
  {https://ui.adsabs.harvard.edu/abs/1999A&A...349.1003T} {349, 1003}

\bibitem[\protect\citeauthoryear{{Varni{\`e}re}, {Tagger}  \&
  {Rodriguez}}{{Varni{\`e}re} et~al.}{2012}]{Varniere2012}
{Varni{\`e}re} P.,  {Tagger} M.,   {Rodriguez} J.,  2012, \mn@doi [\aap]
  {10.1051/0004-6361/201116698}, \href
  {https://ui.adsabs.harvard.edu/abs/2012A&A...545A..40V} {545, A40}

\bibitem[\protect\citeauthoryear{{Watarai} \& {Mineshige}}{{Watarai} \&
  {Mineshige}}{2001}]{Watarai2001}
{Watarai} K.-Y.,  {Mineshige} S.,  2001, \mn@doi [\pasj]
  {10.1093/pasj/53.5.915}, \href
  {https://ui.adsabs.harvard.edu/abs/2001PASJ...53..915W} {53, 915}

\bibitem[\protect\citeauthoryear{{Wijnands}, {Homan}  \& {van der
  Klis}}{{Wijnands} et~al.}{1999}]{Wijnands1999}
{Wijnands} R.,  {Homan} J.,   {van der Klis} M.,  1999, \mn@doi [\apjl]
  {10.1086/312365}, \href
  {https://ui.adsabs.harvard.edu/abs/1999ApJ...526L..33W} {526, L33}

\bibitem[\protect\citeauthoryear{Yan, Ding, Wang, Qu  \& Song}{Yan
  et~al.}{2013}]{Yan2013}
Yan S.-P.,  Ding G.-Q.,  Wang N.,  Qu J.-L.,   Song L.-M.,  2013, \mn@doi
  [Monthly Notices of the Royal Astronomical Society] {10.1093/mnras/stt968},
  434, 59

\bibitem[\protect\citeauthoryear{{van den Eijnden}, {Ingram}  \& {Uttley}}{{van
  den Eijnden} et~al.}{2016}]{vandenEijnden2016}
{van den Eijnden} J.,  {Ingram} A.,   {Uttley} P.,  2016, \mn@doi [\mnras]
  {10.1093/mnras/stw610}, \href
  {https://ui.adsabs.harvard.edu/abs/2016MNRAS.458.3655V} {458, 3655}

\bibitem[\protect\citeauthoryear{{van der Klis}}{{van der
  Klis}}{1989}]{vanderKlis1989}
{van der Klis} M.,  1989, \mn@doi [\araa]
  {10.1146/annurev.aa.27.090189.002505}, \href
  {https://ui.adsabs.harvard.edu/abs/1989ARA&A..27..517V} {27, 517}

\makeatother
\end{thebibliography}

\bsp	 % typesetting comment
\label{lastpage}
\end{document}